\documentclass[prb,aps,floatfix,amsmath,amssymb,superscriptaddress]{revtex4}
\usepackage{graphicx}
\usepackage{epstopdf}
\newcommand{\be}{\begin{equation}}
\newcommand{\ee}{\end{equation}}
\usepackage{amsmath}
\usepackage{amsfonts}
\usepackage{amssymb}
\usepackage{color}

\usepackage{amsthm}
\newtheorem{lemma}{Lemma}
\newtheorem{definition}{Definition}
\newtheorem{theorem}{Theorem}
\newtheorem{conjecture}{Conjecture}

\begin{document}
\title{Trivial Low Energy States for Commuting Hamiltonians, and the Quantum PCP Conjecture}
\author{Matthew B. Hastings}
\affiliation{Duke University, Department of Physics, Durham, NC, 27708}
\affiliation{Microsoft Research, Station Q, CNSI Building, University of California, Santa Barbara, CA, 93106}

\begin{abstract} We consider the entanglement properties of ground states of Hamiltonians which are sums of commuting projectors (we call these commuting projector Hamiltonians), in particular whether or not they have ``trivial" ground states, where a state is trivial if it is constructed by a local quantum circuit of bounded depth and range acting on a product state.  It is known that Hamiltonians such as the toric code only have nontrivial ground states in two dimensions.
Conversely, commuting projector Hamiltonians which are sums of two-body interactions have trivial ground states\cite{bv}.  Using a coarse-graining procedure, this implies that any such Hamiltonian with bounded range interactions in one dimension has a trivial ground state.  In this paper, we further explore the question of which Hamiltonians have trivial ground states.

We define an ``interaction complex" for a Hamiltonian, which generalizes the notion of interaction graph and we show that if the interaction complex can be continuously mapped to a $1$-complex using a map with bounded diameter of pre-images then the Hamiltonian has a trivial ground state assuming one technical condition on the Hamiltonians holds (this condition holds for all stabilizer Hamiltonians, and we additionally prove the result for all Hamiltonians under one assumption on the $1$-complex).
While this includes the cases considered by Ref.~\onlinecite{bv}, we show that it also includes a larger class of Hamiltonians whose interaction complexes cannot be coarse-grained into the case of Ref.~\onlinecite{bv} but still can be mapped continuously to a $1$-complex.

One motivation for this study is an approach to the quantum PCP conjecture.  We note that many commonly studied interaction complexes can be mapped to a $1$-complex after removing a small fraction of sites.  For commuting projector Hamiltonians on such complexes, in order to find low energy trivial states for the original Hamiltonian, it would suffice to find trivial ground states for the Hamiltonian with those sites removed.  Such trivial states can act as a classical witness to the existence of a low energy state.  While this result applies for commuting Hamiltonians and does not necessarily apply to other Hamiltonians, it suggests that to prove a quantum PCP conjecture for commuting Hamiltonians, it is worth investigating interaction complexes which cannot be mapped to $1$-complexes after removing a small fraction of points.  We define this more precisely below; in some sense this generalizes the notion of an expander graph.  Surprisingly, such complexes do exist as will be shown elsewhere\cite{fh}, and have useful properties in quantum coding theory.
\end{abstract}
\maketitle
In this paper, we are interested in the entanglement properties of low energy states of lattice Hamiltonians that are sums of commuting projectors where each projector acts on a small
number of sites.  An example of such a model is the toric code\cite{tc} where each projector acts on at most $4$ sites (in the toric code model, the degrees of freedom are often regarded as sitting on bonds of a square lattice, but throughout this paper we use the term ``sites" to refer to the degrees of freedom).  More
complicated examples include the Levin-Wen\cite{lw} models where projectors act on a larger number of sites.
Further, we are interested in the case in which each site is acted on by a small number of projectors (for example, in the toric code model each site is acted on by $4$ projectors).

We can use such a Hamiltonian to define an interaction graph and a metric: let the sites represent vertices of the graph, draw an edge
between any two sites if there is some projector that acts on both sites, and use the shortest path metric on this graph (if two sites appear in multiple different projectors, we still join them with only one edge).
Then, we have {\it local} interactions on this graph, called the ``interaction graph", which has bounded degree, denoted $d$.
However,
the ground state may have nonlocal entanglement properties.
Examples of models with such nonlocal entanglement include
the toric
code\cite{tc}, Levin-Wen\cite{lw} models and other such lattice models, where the ground state $\psi_0$
is topologically ordered.

This topological order can be defined in a number of ways.  For example, one can consider a dependence of the ground state degeneracy upon
the topology of the lattice.  One can also define a state $\psi_0$ to be topologically ordered if there exists another state $\psi_1$ which
is orthogonal to $\psi_0$ such that $\psi_1$ and $\psi_0$ have the same (or, more generally, the same up to exponentially small error)
reduced density matrices on any set of sufficiently small diameter compared to the system size\cite{bhv} (see also the disk axiom\cite{tc,tqc2,disk1,bhm,bh}).

The definition of topological order that we will use is that no ground state of the Hamiltonian can be constructed, even approximately,
by acting on a product state with a local quantum circuit with bounded depth and range\cite{bhv} as defined more precisely below; conversely, states which can be constructed in this fashion will be called ``trivial".
In this paper, we consider also a weaker question: is there a trivial state whose energy is {\it close} to the ground state energy?  There are different possible ways one could define ``close", but
 in this paper, by
``close" we mean that the {\it energy density}, that is the energy divided by the number of sites, is small (we fix the energy of the ground state to equal $0$).

In order to better define the energy density,
let us fix some notation.  We will be interested in how the energy depends as a function of the number of sites $N$, at fixed degree $d$ of the interaction graph.
We will use ``computer science" Big-O notation in this paper.  As specific examples of this, if we say that
a quantity is $O(1)$ is means that it is bounded by an $N$-independent constant for sufficiently large $N$, while a quantity which  is $O(N)$ is bounded by a constant times
$N$ for sufficiently large $N$.  While this notation is very familiar to computer scientists it may be less familiar to physicists and so we mention it here to ensure that our notation is understood (in contrast often in the physics literature a quantity is called $O(N)$ if it is asymptotically greater than some constant times $N$ and smaller than some other constant times $N$, a property which in the computer science literature is instead referred to as being $\Theta(N)$).
When referring to a quantity such as energy density being $O(1)$, we implicitly are referring to a family of Hamiltonians with different $N$.

We also need to fix some notations regarding locality.  We write
\be
\label{Hsum}
H=\sum_Z h_Z,
\ee
where the sum ranges over sets $Z$,
where the $h_Z$ are commuting projectors, where each projector $h_Z$ acts on the sites in the set $Z$.  Further we assume that the ground state has zero energy so that there is a state which minimizes every term $h_Z$ separately (that is, $H$ is frustration-free).
The bound on the degree $d$ of the interaction graph implies that each term $h_Z$ acts on at most $O(1)$ different sites.  We call a system
satisfying such assumptions a ``locally commuting projector Hamiltonian".  If the Hamiltonian is a sum of commuting terms which need not be projectors and if we remove the assumption that there is a state that minimizes every term $h_Z$ separately, then we call such a Hamiltonian a ``locally commuting Hamiltonian" (as we explain below in \ref{reduce}, for every local commuting Hamiltonian $H$, we can define a local commuting projector Hamiltonian $H'$ such that every ground state of $H'$ is a ground state of $H$, so in this paper we focus on the case of local commuting projector Hamiltonians).
Note that for a locally commuting projector Hamiltonian, since each $h_Z$ is a projector, the operator norm of $h_Z$ is bounded by $1$ (indeed, the norm equals $1$ unless $h_Z=0$), and hence
$\Vert H \Vert=O(N)$.

We
define a local quantum circuit to be a circuit of depth $D_{circuit}=O(1)$ constructed from unitary gates, each of which acts on a set of diameter $R_{circuit}=O(1)$, using
the interaction graph above to define the diameter and with all the gates in a given round of the quantum circuit acting on sets which are disjoint from each other.
We call $R_{circuit}$ the ``range" of the circuit; this terminology is not completely standardized in the literature since often the range refers to the product $D_{circuit} R_{circuit}$, but since $D_{circuit}$ and $R_{circuit}$ are both $O(1)$, the product $D_{circuit} R_{circuit}$ is also $O(1)$.
We call a state produced by acting on a product state with a local quantum circuit a ``trivial state". We also allow the use of ancillas in this definition.  That is, we refer to the Hilbert space of the given system as the ``real" Hilbert space and we may tensor in an additional ``ancillary" Hilbert space on each site, define any product state $\psi_{prod}^{real}\otimes \psi_{prod}^{ancilla}$ on this enlarged space, where $\psi_{prod}^{real,ancilla}$ are product states on the real or ancillary spaces, construct any unitary $U$ from a local quantum circuit (with $U$ acting on both real and ancilla spaces), and then consider the
state $U(\psi_{prod}^{real}\otimes \psi_{prod}^{ancilla})$.
There are a few different senses in which one might imagine allowing the use of ancillas.  One sense is that if
\be
\label{pure}
U(\psi_{prod}^{real}\otimes \psi_{prod}^{ancilla})=\psi_{out}^{real} \otimes \psi^{ancilla}
\ee
for some state $\psi^{ancilla}$ on the ancillas, then we say that the output state
$\psi_{out}^{real}$ is a trivial state (one might choose to require that $\psi^{ancilla}$ also be a product state but we do not require this).
Our construction later in subsection \ref{gs1l} will use ancillas in this sense, with the dimension of the ancilla space on each site being $O(1)$.

One might choose instead another sense of allowing the use of ancillas.  We could trace the state $U(\psi_{prod}^{real} \otimes \psi_{prod}^{ancilla})$ over the ancillas to define a density matrix on the real degrees of freedom, and we could then refer to such a density matrix as a trivial density matrix\cite{betatopo}.  Allowing this density matrix to be mixed, as opposed to Eq.~(\ref{pure}) where it is necessarily pure, amounts to enlarging the definition of which states we consider to be trivial.  Our construction later will not require the use of such mixed trivial states.

It is possible to show, by extending arguments such as those in Ref.~\onlinecite{bhv} and Ref.~\onlinecite{leshouches}, that we cannot construct a trivial state whose
energy is at most $O(1)$ above the ground state energy for a family of toric code Hamiltonians of increasing system size $N$ (we will not give a proof of this here).
So, we ask a weaker question: is it possible, for every $\epsilon>0$, to construct a trivial state whose energy is at most $\epsilon N$
above the ground state?   In this case, the depth of the quantum circuit required may depend upon $\epsilon$, but we seek bounds on its
depth which are $N$-independent.  Defining the energy density to be the energy divided by $N$, we are asking whether we can find trivial
states with energy density at most $\epsilon$ above the ground state, for every $\epsilon>0$.

One motivation for this question is for its application to the quantum PCP conjecture\cite{qpcp}.  This conjecture considers Hamiltonians which are a sum of terms as in Eq.~(\ref{Hsum}), though typically one does not require that the terms commute with each other.  One requires that the interaction graph have degree $O(1)$ and that each site have Hilbert space dimension $O(1)$, and each $h_Z$ has norm bounded by $\Vert h_Z \Vert \leq 1$.  Then, roughly speaking, the conjecture is that there is a constant $c>0$ such that  it is QMA-hard to approximate the ground state energy more accurately than $c \Vert H \Vert$.  More precisely, this problem of approximating the ground state energy is formalized by saying that it is QMA-hard to answer the decision problem of whether the energy is less than $E$ given a promise that if not, the ground state energy is greater than $E+c \Vert H \Vert$.  Currently, this conjecture is completely open.

If a Hamiltonian has a trivial low energy state, then this state can be used as a classical witness for the existence of a low energy state because one can efficiently compute the energy density of this state on a classical computer (note that one can still efficiently compute the energy density even if ancillas are allowed and even if the density matrix on the real degrees of freedom is not pure, so long as the ancilla dimension remains bounded).  Thus, in order for the quantum PCP conjecture to be true, it is necessary that there exist families of Hamiltonians $H(N)$, where each Hamiltonian $H(N)$ is defined on a system of $N$ sites, with uniform bounds on the Hilbert space dimension and degree of the interaction graph of such Hamiltonian such that for some $\epsilon>0$, there is no finite $D_{circuit}$ and $R_{circuit}$ such that for all $N$ there is a state of energy density at most $\epsilon$ for $H(N)$ which can be constructed by a quantum circuit with depth $D_{circuit}$ and range $R_{circuit}$ acting on a product state.  Constructing any such family would be very interesting even if it did not prove the quantum PCP conjecture.

In this paper, we present an attack on this question in the specific case of commuting Hamiltonians.  This attack is based on the following idea: in order to find a low energy state for a given Hamiltonian, it suffices to find a zero energy state on a modified Hamiltonian with some small fraction of the interaction terms $h_Z$ removed.
For many Hamiltonians, we will show how to remove certain terms to find a trivial ground state.  Before giving specifics, we motivate with two examples.

Suppose on the one hand that the Hamiltonian described interactions in a finite dimensional system, as the toric code Hamiltonian does (that is, there is an underlying $D$-dimensional lattice such that all terms in the Hamiltonian have bounded range with respect to that lattice).
For simplicity, suppose that the lattice is a hypercube of size $L$ on each side.  In this context, by a ``hypercube", we mean that each site in the lattice is labelled by $D$ different integers, each ranging from $1...L$ so that there are $L^D$ sites, and two sites are neighbors if the first site is labelled by integers $i_1,...,i_D$ and the second is labelled by integers $j_1,...,j_D$ such there is some $b$ in the range $1...D$ such that $i_a=j_a$ for $a \neq b$ and $i_b=j_b \pm 1$.  While this definition of a hypercubic lattice is standard for physicists, we include this definition here because in the computer science literature the term ``hypercube graph" has a very different definition.
For such a lattice, it is easy to construct the desired trivial state for any $\epsilon>0$.  Break the $D$-dimensional lattice up into small hypercubes of size $l$ on each side, for some $l$ ($l$ will depend upon the desired $\epsilon$ but will not depend upon $L$).  Now, define $H'$ to be the Hamiltonian which is the sum of $h_Z$ over all sets $Z$ such that all sites in $Z$ are in the
same small hypercube.  That is: drop all terms on the boundary of a small hypercube which connect that small hypercube to another small hypercube.  The ground state
of $H'$ is a product state on the small hypercubes, and hence it is a trivial state since it can be obtained by acting on a product state on the original lattice with a quantum circuit of depth $1$ with unitaries acting on sets of diameter $l$.  Further, the norm $\Vert H'-H \Vert$ is
bounded by a constant times $N/l$.  To see this, realize that the lattice has $L^D$ sites, but the total number of sites on the boundaries between the small hypercubes is of order $L^{D}/l$.  Thus, by choosing $l$ of order $1/\epsilon$, the ground state of $H'$ provides an example of a trivial state which has energy density for the Hamiltonian $H$ which is within $\epsilon$ of the ground state energy density of $H$.
As a side remark, note that if $l$ does not exactly divide $L$ then this is not a problem: we can simply ``pad" the lattice by adding small number of sites to increase $L$ to make $l$ divide $L$.
As another side remark, not needed for the rest of the paper, note that it is also interesting in this problem to consider the case where the $h_Z$ need not be projectors so that the  norms of the terms $h_Z$ are no longer bounded by unity; in this case, one can show that there exists some way of breaking the lattice up into smaller hypercubes where the terms that are dropped have norm at most $\epsilon \sum_Z \Vert h_Z \Vert$.  However, for the rest of the paper we do not consider this problem of varying norms of the terms.

This first example is worth bearing in mind: it means that the approximation problem for that finite dimensional system can be solved in a time which is {\it linear} in $N$, although the time required scales exponentially with $1/\epsilon^d$ (indeed, note that we can {\it find} the desired trivial state by diagonalizing the Hamiltonian on each hypercube; glossing over certain details in the time required to do floating point arithmetic, this takes a time exponential in $1/\epsilon^d$,  so the problem is in $P$).  This first example means that we must instead turn
to families of Hamiltonians which are not defined on finite dimensional lattices in order to try to find a family which does not have a trivial state with arbitrarily small energy density for all $N$.  

The natural next thing to consider is Hamiltonians where the interaction graph is an expander graph.  These are graphs for which given any
set of sites $X$, such that the cardinality of $X$ is sufficiently small compared to $N$, the number of neighbors of the set $X$ is lower
bounded by a constant times the cardinality of $X$.  Such graphs prevent the kind of argument we used above: that argument was based on dividing the system up into small hypercubes, such that the surface-to-volume ratio (the number of terms that connected the hypercube to other hypercubes, divided by the number of sites in the hypercube) became small as $l$ became large.  Such graphs play a large role in Dinur's proof\cite{dinur} of the classical PCP theorem.

Many examples of expander graphs are high girth.
Suppose in fact the graph has no triangles.
Then, all of the projectors act on at most two different sites.
If all of the projectors in the Hamiltonian act on at most two sites, it has been shown by Bravyi and Vyalyi\cite{bv} that  the problem of finding
the ground state is in NP, using $C^*$-algebraic considerations. 
There is a ground state on such a graph which has very simple entanglement properties, so that entanglement is only between nearest neighbor sites.
We will discuss their result in more detail in section \ref{solvesouse} since we will make use of some of the same techniques.
Such ground states will be trivial by our definition.

So, one might next turn to expander graphs where the interaction graph contains triangles or larger cliques.  For many such graphs, though, it is still possible to cluster the sites in certain ways, using clusters of size $O(1)$,
defining a coarse-grained interaction graph such that Ref.~\onlinecite{bv} still applies.  As a simple example of such a graph, take an expander graph $G$ without triangles, and define a new graph $G'$ which has two vertices $v_1,v_2$ for every vertex $v$ in $G$; define edges in $G'$ between $v_1$ and $v_2$ for all $v$ and define edges between $v_a$ and $w_b$ for all $a,b$ whenever $G$ has an edge between $v$ and $w$; this graph $G'$ has triangles but one can cluster $G'$ using clusters of size $2$ by combining $v_1$ and $v_2$ into a single vertex for each $v$ to obtain a graph with no triangles: the resulting graph in fact is precisely the original graph $G$.
In this case, again, the ground state is a trivial state.

However, this still does not exhaust the class of graphs for which we can find a trivial ground state.
We construct a family of graphs of increasing $N$ and fixed $d$ which are generated by taking certain random high girth graphs and taking interactions that involve triples of sites all within distance $2$ of each other; such an interaction graph is just a power of a random high girth graph and still itself has high girth.  However, we show that the resulting interaction graphs in such a family, with high probability, cannot be coarse-grained into a triangle-free graph using clusters of size $O(1)$ (recall that to say that something is not possible with high probability means that the probability that it is possible goes to zero as $N\rightarrow \infty$).  In fact, we show that for
sufficiently small $\epsilon$ with high probability one cannot
remove a fraction $\epsilon$ of vertices (also removing their attached edges) and then coarse-grain the graph into a triangle-free graph using clusters whose size is independent of $N$.
 More surprisingly, the graphs in this family can all be locally coarse-grained into a tree as described later, meaning that for any vertex in the graph, one can take the subgraph containing vertices within some distance of that vertex (this distance diverges as $N \rightarrow \infty$) and coarse-grain that subgraph into a tree using clusters of size $O(1)$, and yet the ability to perform such local coarse-graining does not imply the to perform global coarse-graining into a triangle-free graph.  
However, although these graphs cannot be coarse-grained into a triangle-free graph, we will show that Hamiltonians whose interaction graph is such a graph do indeed have a trivial ground state. To do this, we will find it useful to generalize the idea of interaction graphs to ``interaction complexes", defining simplicial $2$-complexes to describe the support of interactions in a system.  For a class of complexes, that we call $1$-localizable with range $R$, we show (under one technical assumption on the Hamiltonian that holds for {\it all} stabilizer Hamiltonians with such an interaction complex) how to find an exact ground state using a bounded depth and range quantum circuit, with the depth and range of the quantum circuit depending only upon $d$ and $R$, and not $N$.  This class includes, as we said, these powers of high girth graphs (for such powers, the technical assumption on the Hamiltonian is {\it not} needed), but also includes other graphs as well.

The paper is organized as follows.  We begin by considering the question of coarse-graining graphs into triangle-free graphs, in an attempt to apply the result of Ref.~\onlinecite{bv}.  After all, if it were always possible to either coarse-grain a graph into a triangle-free graph, or at least to delete a small fraction of edges and then perform such coarse-graining, then we would have no need to define the more general class of $1$-localizable complexes.  We construct the family of random graphs discussed above to show that most such graphs, even though they can be locally coarse-grained into a tree, cannot be coarse-grained into a triangle-free graph, even after deleting a small fraction of vertices.
We then define interaction complexes, and define $1$-localizable complexes as those that can be continuously mapped to a $1$-complex such that the pre-image of any point in the $1$-complex has bounded diameter.
We then show how to construct a trivial ground state for Hamiltonians with $1$-localizable interaction complexes under either a technical assumption on the Hamiltonian or under the assumption that the $1$-complex has large enough girth (we conjecture that neither of these assumptions is necessary, but we have not been able to prove that).

The next part of the paper raises the question of whether every interaction complex can be turned into a $1$-localizable interaction complex by removing a small fraction of cells (removing a term $h_Z$ from the Hamiltonian will remove certain cells from the complex).  In this paragraph, we give definitions which
allow us to give a formalization this question.   We define ``$0$-hyperfinite" and ``$1$-hyperfinite" families of complexes and the formalization of this question is ``are all families of complexes with uniform bounds on their local geometry (i.e., uniform bounds on the number of cells attached to any given cell) $1$-hyperfinite?".  The family of complexes which are not $0$-hyperfinite is closely related to expander graphs, though the definition is slightly different; instead, $0$-hyperfinite complexes are the same as hyperfinite\cite{hyperfinite} graphs.  In a separate work\cite{fh}, it will be shown that families of complexes with uniform bounds on local geometry which are not $1$-hyperfinite exist and we will give an explicit construction of these complexes; these complexes serve as a natural place to look in trying to prove the quantum PCP conjecture.
Finally, in an appendix we discuss the relationship between $1$-localizable complexes and properties of the cover of the complex, and comment on the relation between this approach and quantum belief propagation\cite{qbp1,qbp2,qbp3}.

Before beginning, a comment on the toric code in two dimensions.  As outlined above, we can find an approximate ground state by breaking the problem up into square of linear size $l$, getting an energy density of order $1/l$.  Suppose instead we puncture the square lattice by removing interactions on certain holes, with the holes being spaced on a square lattice of larger linear size $l$ on each side.  Such a Hamiltonian will have a $1$-localizable complex, and in fact an exact ground state can be constructed with a quantum circuit of range of order $l$ and depth of order unity\cite{betatopo}; in this case, the energy density is of order $1/l^2$.  Thus, this approach also gives one a lower energy density than the simple approach of breaking the lattice into hypercubes.  This approach can be carried out in higher dimensions.  For example, in three dimensions, one must remove interaction terms lying on lines; in general one must remove interaction terms on sets of co-dimension $2$. 

\section{Different Graphs}
\label{countersubsec}
The method of Ref.~\onlinecite{bv} is applicable to interaction graphs without triangles.  More generally, they consider Hamiltonians in which each term acts on at most two sites, which includes all interactions graphs without triangles, since any term acting on three or more sites induces a triangle in the interaction graph, but for our purposes in this section, let us assume that wherever there is an interaction graph containing a triangle, then there might be a term in the Hamiltonian acting on all three sites and see how far we can go using Ref.~\onlinecite{bv}.  One result of this section will be to show that for a large class of families of graphs we can coarse-grain the graph, using clusters of size $O(1)$, to obtain a graph which is triangle-free. Another result will be the existence of certain families graphs where, for any given site, one can use clusters of size $O(1)$ to obtain a coarse-grained graph which is free of triangles for a large radius away from that site, but for which one {\it cannot} use clusters of size $O(1)$ to obtain a coarse-grained graph which is globally free of triangles, even if one is allowed to remove a small fraction of edges; this will motivate our interest in defining $1$-localizable complexes later to generalize the class of systems that we can solve exactly.

Some preliminary notation:
by coarse-graining a graph, we mean the following: 
\begin{definition}
Given a graph $G$, we define a coarse-grained graph $G'$ as follows.  Let $C_1,C_2...$ be sets of vertices of $G$.  We called these ``clusters".  Let these clusters be disjoint, and let each vertex of $G$ be in one of the clusters.  Then, the coarse-grained graph $G'$ has one vertex corresponding to each cluster, and there is an edge $(i,j)$ in $G'$ if and only if there is a vertex $v \in C_i$ and a vertex $w\in C_j$ such that the edge $(v,w)$ is in $G$.
\end{definition}
If the graph $G$ is infinite, then the set of clusters $\{C_i\}$ may also be infinite.

We define the following transformation on graphs.  If $G$ is a graph, then the $R$-th power of $G$, written $G^R$, is 
the graph containing the same vertices as $G$, and with an edge $(i,j)$ in $G^R$ between vertices $i$ and $j$ whenever the distance between $i$ and $j$ in $G$ is at most $R$, for some $R$.  This transformation $G^R$ is a simple way to take a graph without triangles and construct a new graph containing triangles, which will be useful for our examples now.
Finally, we say that a graph $G'$  is constructed by ``removing at most a fraction $\epsilon$ of the edges of $G$" if $G'$ has the same vertices as $G$, and the set of edges in $G'$ includes all but a fraction $\epsilon$ of the edges of $G$.

\begin{figure}
\includegraphics[width=1.9in]{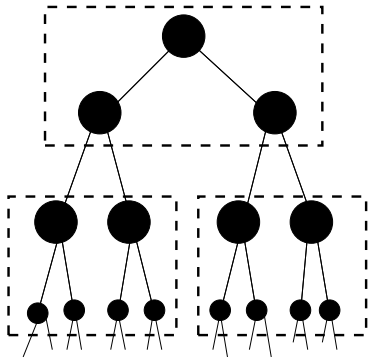}
\caption{Illustration of coarse-graining.  Dashed lines represent clusters.  Here the graph shown $T$ is a binary tree, and the clusters coarse-grain $T^2$ to a tree.  Only part of the process is shown.}
\label{coarsetree}
\end{figure}

Consider a graph $T$ which is a binary tree (or any other tree with degree $O(1)$).  
Then, for any $R$ which is  $O(1)$, $T^R$ can be clustered into clusters of size $O(1)$ such that the resulting coarse-grained graph is a tree.
See Fig.~\ref{coarsetree} for an illustration for a binary tree with $R=2$ (we omit a general proof of this statement, since the proof in the general case is a natural generalization of the procedure in the figure).

Given that we have shown that it is possible to coarse-grain the $R$-th power of a tree graph to produce a tree graph, using clusters of size $O(1)$, it would be natural to conjecture that something similar holds for high girth graphs.  Namely, we would like to conjecture that:
\begin{conjecture}
{\bf This conjecture is false but natural!} For all $d,R$ there is an $r$ such that the following holds.
Let $E$ be any graph with girth at least $r$ and degree at most $d$.    Then, it is possible to coarse-grain $E^R$ using clusters of size $O(1)$ (the size may depend upon $d,R$ but not upon the size of $E$) to obtain a graph with no triangles.
\end{conjecture}
In fact, given that we are interested in this paper in removing small fractions of vertices from graphs, we might at least hope for the weaker conjecture that
\begin{conjecture}
{\bf This conjecture is also false but natural!} For all $d,R$ and for all $\epsilon>0$ there is an $r$ such that the following holds.
    Let $E$ be any graph with girth at least $r$ and degree at most $d$. 
Then, it is possible to define a graph $G$ by removing
at most a fraction $\epsilon$ of the edges of $E^R$, and then coarse-grain $G$ using clusters of size at most $O(1)$ (the size may depend upon $d,\epsilon$ but not upon the size of $E$) to obtain a graph with no triangles.
\end{conjecture}

The counter-example to these conjectures is to choose a random, high girth graph for $E$ from a certain ensemble of random graphs that we now describe, and then set $R=2$.  We construct $E$ by first iteratively constructing a random graph $E_0$ with $N$ vertices as follows.
We begin with a graph with no edges.  Then, for each vertex $i$, we choose $d/4$ random other vertices, and add edges to $E_0$ connecting those pairs of vertices, so the average degree of
the graph is $d/2-O(1/N)$ (the $O(1/N)$ correction is there because it is possible that the same edge may be added twice by this procedure; if this procedure adds an edge two or more times, we continue to have only one edge connecting those two vertices).
This construction of $E_0$ can be described in pseudo-code as follows. Initialize the $N$-by-$N$ adjacency matrix to $0$.  Then, use the following
 algorithm:
\begin{itemize}
\item[1.] {\bf for} $i=1...N$
 \begin{itemize}
\item[2.]      {\bf for} $j=1...d/4$
\begin{itemize}
 \item[3.] choose a random $k$ in the range $1...N$ such that $i \neq k$.  If the $i,k$ entry of the adjacency matrix is equal to $0$, then set it to $1$.  Do the same to the $k,i$ entry so that the matrix is symmetric.
\end{itemize}
 \item[4.]     {\bf end for}
\end{itemize}
 \item[5.] {\bf end for}
\end{itemize}

Having constructed $E_0$, we then
remove any vertex in that graph participating in a loop of length $l\leq 2r$, and also remove any vertex with more than $d$ edges connecting to it and let
the resulting graph be $E$, giving a graph $E$ of girth at least $r$ and which has maximum degree $d$ (some vertices may have degree less than $d$).
We follow this procedure for constructing our random graph to guarantee that certain random events are independent; other procedures, such as choosing simply a random graph of fixed degree, are expected to work also and might give better estimates, but would require more care in dealing with correlations between events.

We claim that typically only $O(1)$ vertices need to be removed from $E_0$ to get rid of loops of length at most $l$ for $l$ which is $O(1)$.  To see this, estimate the probability that there is a loop of length $l$ starting from a given vertex.  This probability
 is exponentially large in $l$, but is of order $1/N$, where $N$ is the number of vertices in the graph.  Thus, summing over vertices, the number of such loops
is bounded by a quantity that is $N$-independent (but exponentially large in $l$).  Removing all vertices that participate in a loop of length $l\leq 2r$ creates a graph that has girth at least $r$.  Let $N_r$ be the number of edges removed when removing vertices with degree more than $d$.  With probability approaching $1$ for large $N$, this number $N_r$ is bounded by $N$ times a constant which is exponentially small in $d$.

We now show that
\begin{theorem}
For any sufficiently small $\epsilon>0$ and any sufficiently large $d$, for any $r$, it is not possible, with high probability, to find a clustering of $G$ (where $G$ is the graph $E^2$ with at most a fraction $\epsilon$ edges removed) using clusters of size at most $O(1)$ to obtain a triangle-free graph.
\begin{proof}
The proof is probabilistic: we upper bound the number of possible clusterings of $N$ vertices with clusters of size at most $O(1)$ as well as the probability that for a random graph $E_0$ a random clustering will give a coarse-grained graph $E_0^2$ with fewer than $cN$ triangles for some $c>0$.  Since the vertex set of $E$ is a subset of the vertex set of $E_0$, every clustering of $E_0$ induces a clustering of $E$: given a clustering of $E_0$ using sets $C_i$, we define a clustering of $E$ using sets $C_i \cap V$, where $V$ is the vertex set of $E$.
However, $E^2$ may have fewer triangles for a given cluster than $E_0^2$, since we have removed some edges.
However, since we will have shown that typically such a coarse-grained graph has at least $cN$ triangles, this will imply that, for sufficiently large $d$ and sufficiently small $\epsilon$, it is not possible to get rid of all the triangles by first removing those $O(1)$ vertices which are in loops, then removing $N_r$ edges (since $N_r$ is exponentially small in $d$, this removes a fraction of triangles which is exponentially small in $d$), and then removing at most a fraction $\epsilon$ of the edges, implying that $E^2$ still has triangles.

To upper bound the number of clusters, note that we can specify a clustering as follows.  Let $C$ be the maximum size of a cluster.  There are at most $N$ clusters in the graph.  We specify a clustering by first listing how many vertices are in cluster $C_1$, then how many vertices are in cluster $C_2$, and so on up to the last cluster.  Then, if there are fewer than $N$ clusters, we pad the list with zeroes to obtain a list of $N$ numbers.  Then we list which vertices are in $C_1$, then list which vertices are in $C_2$, and so on.  The first list consists of at most $N$ different numbers, ranging from $0$ to $C$, so the number of choices there is bounded by $(C+1)^N$.  The second list is a permutation, so the number of choices is bounded by $N!\leq N^N$.  So, the number of clusterings is bounded by $((C+1)N)^N$.

To upper bound the average, over random graphs $E_0$, of probability that a randomly chosen cluster will give a coarse-grained graph without triangles, we work in reverse: we upper bound the probability, for any given clustering that a randomly chosen graph $E$ will lead to a coarse-grained graph with fewer than $cN$ triangles.  Consider a given cluster $C_i$.  This cluster has $|C_i|$ vertices.  We say that a vertex in $C_i$ is ``in the center of the triangle" if one of the $d/4$ edges added for that vertex connects that vertex to another cluster $C_j$ and another one of those $d/4$ edges connects that vertex to a cluster $C_k$ with $i,j,k$ all different.
Thus, it is the center of a triangle unless
all of those edges either go to a vertex in $C_i$ or to a vertex in cluster $C_j$ for some $j\neq i$.  For any given $i$, any given vertex 
in $C_i$ and any given $j$, the probability that all of these edges added for that vertex connect to $C_i$ or $C_j$ is at most $((|C_i|+|C_j|)/N)^{d/4}\leq (2C/N)^{d/4}$.  There are at most $N$ possible choices of $j$, so the probability that a vertex is not the center of a triangle is at most $N (2C/N)^{d/4}$.  With this slightly complicated construction of a graph, the probabilities of vertices being centers of triangles are all independent.  So, we can estimate the probability that at most $cN$ vertices are centers of triangles (and hence that at least $N-cN$ vertices are not centers of triangles)
as being at most
\be
\sum_{M=0}^{cN} {N \choose M} \Bigl(N (2C/N)^{d/4}\Bigr)^{N-M} 
\ee
where we used the fact that the number of ways to have $M$ vertices being centers of triangles is ${N \choose M}$.
Suppose that $d>4$.  Then, for sufficiently large $N$, the quantity
$N (2C/N)^{d/4}$ is less than $1$ and the largest term in the sum over $M$ is for $M=cN$, so we can bound the result by
$N^{cN} \Bigl(N (2C/N)^{d/4} \Bigr)^{(1-c)N}$ for large enough $N$.

Multiplying this by the number of clusterings, which is at most $((C+1)N)^N$, the probability that a randomly chosen graph has a clustering with fewer than $cN$ triangles goes to zero for large $N$ (at fixed $C$) so long as $-(d/4-1)(1-c)+c+1<0$ as one may see by counting powers of $N^N$.  So, for $d>8$, for sufficiently small $c$, this probability goes to zero, completing the proof.
\end{proof}
\end{theorem}

Thus, we can construct families of graphs such that the neighborhood  which can be clustered, using clusters of size $O(1)$, into graphs which are trees up to any desired distance near any given vertex, but such that it is not possible to clusters these graphs globally into triangle-free graphs.  Thus, the method of Ref.~\onlinecite{bv} does  not work to solve Hamiltonians with such graphs as the interaction
graph, motivating the construction of section \ref{solvesouse}, in which we provide a method which can handle such Hamiltonians.

\section{Interaction Complexes}
\subsection{Definition of Interaction Complex}
We now define ``interaction complexes", which generalize the idea of an interaction graph.

A graph may be regarded as a simplicial $1$-complex, with the edges corresponding to $1$-cells and the vertices corresponding to $0$-cells.
Each $k$-cell is a $k$-dimensional subset of Euclidean space.
We refer to this as an ``interaction $1$-complex" or interaction graph.

To define an interaction $2$-complex for a Hamiltonian $H$, we define a $0$-cell for every site.  For every term $h_Z$ in the Hamiltonian, for every pair of sites $i,j\in Z$ with $i\neq j$, we attach a $1$-cell to the $0$-cells corresponding to that pair of sites.  We identify all $1$-cells connecting the same pair of $0$-cells.
Given three sites $i,j,k$ which are all in some set $Z$ for which a term $h_Z$ appears in the sum in Eq.~(\ref{Hsum}), attach a $2$-cell to the three $1$-cells whose faces are the $0$-cells corresponding to
those three sites, and again identify all $2$-cells attached to the same three $1$-cells.  One may continue in this fashion and define an interaction $k$-complex, by attaching, for all $l\leq k$, an $l$-cell whenever $l$ sites appear in the same set $Z$ for some $h_Z$.  However, in this paper we will only be interest in interaction $2$-complexes, and so from now on we simply use the term ``interaction complex" to refer to the interaction $2$-complex.  When we refer to the ``degree" $d$ of an interaction complex, we mean the degree of the graph that is the $1$-skeleton of the complex.

We have used the term ``an interaction complex" rather than ``the interaction complex" for a reason.  For a given Hamiltonian, there might be several different ways of writing it as a sum of commuting terms $h_Z$.  Hence, one might define different interactions complexes for the same Hamiltonian.  Given any interaction complex for a Hamiltonian, any other complex which contains that first complex as a subcomplex is also an interaction complex for that Hamiltonian.

We place a metric on the complex, by defining each edge to have length $1$, and using the shortest path metric.  This metric reproduces the usual graph metric between vertices on the graph (the distance between neighboring vertices is equal to $1$).  We extend this metric by continuity to the $2$-cells.  We choose the metric so that every point in a $2$-cell is distance at most $1/2$ from every $0$-cell.

Note that defining the interaction complex allows some further flexibility in describing interactions than the interaction graph does.  Suppose a triangle connects three sites $i,j,k$ in the interaction graph.  Then, there may or may not be a $2$-cell in the interaction complex involving those three sites, depending upon whether or not all three sites appear in the same interaction term.  
That is, given an interaction graph $K_1$ for some Hamiltonian $H$, the complex obtained by attaching a $2$-cell to every triangle in $K_1$ is always an interaction complex for that Hamiltonian, but, depending upon the support of the interaction in $H$, there may also exist interaction complexes for $H$ with not all such $2$-cells attached.

\subsection{$k$-Localizable Complexes}
Here we define a class of interaction complexes that we call $1$-localizable with range $R$.
This includes the class of complexes obtained by attaching $2$-cells to graphs which can be coarse-grained into triangle-free graphs using clusters of diameter at most $R$.  However, it generalizes this case, because one can show that with high probability the interaction complexes obtained by attaching $2$-cells to the graphs constructed in section \ref{countersubsec} are also $1$-localizable even though these graphs cannot be coarse-grained to triangle free graphs.
In general, if $G$ is a graph with girth larger than $3R$ then the complex obtained from $G^R$ by attaching $2$-cells to all triangles is a $1$-localizable complex with range $R$.  
Further, we will show in subsection \ref{gs1l} that,
using methods building on Ref.~\onlinecite{bv}, if a commuting projector Hamiltonian has a $1$-localizable interaction complex then it has a trivial ground state under one of two assumptions (either a technical condition or an assumption on the girth of the complex as discussed later).

We define $k$-localizable complexes by:
\begin{definition}
\label{defn1}
A metrized simplicial $l$-complex $K_l$ is  ``$k$-localizable with range $R$" if there exists a continuous function $f$ from $K_l$ to some metrized simplicial $k$ complex $K_k$ (using the same graph metric as above on the edges of $K_k$, extended as above by continuity to the higher cells of $K_k$) such that the diameter of the pre-image of any point in $K_k$ is bounded by $R$.
\end{definition}

We now sketch the claim that the complex obtained by attaching $2$-cells to the triangles of $G^R$ for high girth $G$ is $1$-localizable, first considering the case $R=2$.  Some of the triangles involves three vertices, $i,j,k$ with $j$ neighboring $i$ and $k$ in $G$.  For any such triangle, the edge from $i$ to $k$ in $G^R$ is free, meaning that it appears in no other triangle (this holds because the girth of $G$ is large enough).  
So, we can map the midpoint of this edge to the $0$-cell corresponding to $j$, and map the edge onto the $1$-cells corresponding to the edge $i,j$ and edge $j,k$ (map each half of the edge onto one such $1$-cell).  
Doing this for each triangle gives the needed map.  Note that there are also triangles $i,j,k$ where $i,j,k$ all are neighbors of some fourth vertex $l$; however, the map described above also maps these triangles onto edges of the original graph.
A similar procedure works for arbitrary $R$ whenever the girth is greater than $3R$.  Consider each triangle with three vertices $i,j,k$ with distance $R$ from $i$ to $k$.  This edge is free as the girth is large enough, and so we can map the $2$-cell corresponding to this triangle onto the union of the two $1$-cells corresponding to the other edges.  We do this for each such triangle.  We then repeat for each triangle with distance $R-1$ between $i$ and $k$, and so on.

\begin{figure}
\includegraphics[width=3in]{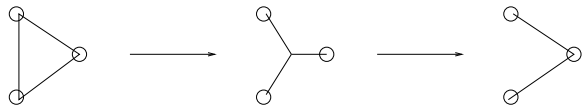}
\caption{Illustration of map $f$.  The left-most image shows an interaction complex $K_2$, with circles being $0$-cells and edges being $1$-cells.  Assume that there is a $2$-cell attached to the triangle.  The middle image shows a map to a $1$-complex with four $0$-cells; three of the $0$-cells are shown as open circles and are images of a $0$-cell in $K_2$, while the fourth $0$-cell is the image of some points in the interior of the $2$-cell as well as some points in the interior of the $1$-cells in $K_2$.  The right-most image shows a way to deform the map further so that every $0$-cell in $K_1$ is the image of some $0$-cell in $K_2$.}
\label{good}
\end{figure}

Before proceeding, it is worth recalling the concept of a simplicial map and of a simplicial approximation.  Using this idea allows us to avoid many subtleties of continuous functions and deal with more combinatoric questions.  A simplicial map is a map from one simplicial complex to another such that the images of the vertices of a simplex span a simplex.  In particular, the image of a $0$-cell is a $0$-cell and a simplicial map is completely determined by its action on $0$-cells.  As an example, let $K_2$ be the $2$-complex obtained by attaching $2$-cells to the triangles of $E^2$ for one of the graphs $E$ considered previously.  Then, a simplicial map from $K_2$ to a $1$-complex $K_1$ would define a clustering of the vertices of $E^2$ such that the coarse-grained graph is triangle-free: each cluster is the set of $0$-cells in the pre-image of a given $0$-cell in $K_1$.  As we have shown, no such simplicial map exists with small diameter of pre-images for the given graphs $E$.
However, by subdividing the simplices of $K_2$ and of $K_1$, it is always possible to approximate (up to a slight deformation) a continuous map by a simplicial map.  This is a fundamental theorem in topology called the simplicial approximation theorem.  Since the error in the approximation can be made arbitrarily small, this approximation has no effect on the notion of ``$1$-localizability", and even the value of the range $R$ is unchanged.
Consider, for example, the map of Fig.~\ref{good}.  This map from the left image to the middle is not simplicial; however, one can subdivide the $2$-cell on the left into four $2$-cells, subdividing each $1$-cell on the left into two $1$-cells, so that the map becomes a simplicial map.

Also, to avoid subtleties, we will assume in this paper that all complexes that we consider are locally finite.  Here, a complex $K$ is ``locally finite" if every $0$-cell is attached to a  finite number of $1$-cells and every $1$-cell is attached to a finite number of $2$-cells.  We will, at certain points later in the paper when discussing covers, consider complexes with an infinite number of $0$-cells and $1$-cells, but always we will assume that the complex is locally finite.
One reason for the choice of locally finite complexes is that it will be used in the proof of the next lemma to show that a certan process terminates.

Suppose $K_2$ is a $1$-localizable complex.
We will show that we can assume in definition \ref{defn1} that, at the cost of a slight increase in $R$ we can show that $f$ has certain
useful properties which we here define.  
\begin{definition}
Suppose that the image of every $0$-cell in $K_2$ is a $0$-cell.
Suppose for every $2$-cell we can pick some point in the interior of the $2$-cell which we call the center of the $2$-cell such that every $0$-cell in $K_1$ is either the image of some $0$-cell in $K_2$ or is the image of the center of a $2$-cell.  Finally, suppose that the inverse image of every point in $K_1$ is a path connected set in $K_2$.  In this case we say that $f$ is {\bf good}.
\end{definition}
As an example, consider the map of Fig.~\ref{good}.  The map from $K_2$ on the left of the image to $K_1$ in the center of the image is good since we can choose the center of the $2$-cell to be some point in the inverse image of the $0$-cell in the center of the middle image.  To explain this definition, we note that the point of choosing the ``center" of the $2$-cell is that we allow only one such point for each $2$-cell; we will see that the point of this is to bound the number of $0$-cell in $K_1$: every such $0$-cell will be the image of a $0$-cell in $K_2$, except for at most one extra $0$-cell per plaquette in $K_2$.

The point of choosing ``good" $f$ is that once we have made this choice, a certain relation will be implied between the distance in $K_2$ and $K_1$.  It will 
turn out  (as we also show below) that there is a bound $l_{max}$ on the length (as defined below) of the image of every $1$-cell in $K_2$; this bound will be expressed solely in terms of $d$ and $R$.  Conversely, there will also be a bound on the diameter of the pre-image of every $1$-cell in $K_1$; this bound again will be solely expressed in terms of $d,R$.  That is, once we have chosen a good $f$, there will be some relation between the metrics on $K_2$ and $K_1$ and, in a sense, there will be a bound on the distortion that the map $f$ introduces: given any set $S_1$ in $K_1$, we can bound
\be
\label{bddist1}
{\rm diam}(f^{-1}(S_1)) \leq {\rm const}\times {\rm diam}(S_1) + {\rm const.},
\ee
and given any set $S_2$ in $K_2$ we can bound
\be
\label{bddist2}
{\rm diam}(f(S_2)) \leq {\rm const}\times {\rm diam}(S_2) + {\rm const.},
\ee
where in both equations the constants depend only upon $d,R$.  
Finally, it will turn out that the degree of $K_1$ (that is, the maximum number of $1$-cells attached to any $0$-cell) is bounded by a function of $d,R$.

Before showing how to construct a good $f$, we note that the above bounds (\ref{bddist1},\ref{bddist2}) would not necessarily have held if we had not chosen $f$ to be good.
For example, to see how Eq.~(\ref{bddist2}) can be violated, consider a complex $K_2$ with two $0$-cells, $i,j$, and one $1$-cell connecting them.  Choose any integer $n\geq 1$, and define a new $1$-complex with a total of $n+1$ $0$-cells, called $c_0,c_1,c_2,c_3,...,c_n$, and with $n$ $1$-cells, called $e_0,e_2,...,e_{n-1}$, where $e_a$ attaches to $0$-cells $c_a$ and $c_{a+1}$.  That is, the $1$-complex is just a line graph.  Map $i$ to $c_0$ and map $j$ to $c_n$.  For integer $k$, map the point in $K_2$ a distance $k/n$ from $i$ onto the $0$-cell $c_k$.  That is, we ``subdivide" the line to obtain $K_1$ from $K_2$ by adding $n-2$ additional $0$-cells.  The distance between $i$ and $j$ is $1$ but the distance between their images is $n$, which can be arbitrarily large.
To see how Eq.~(\ref{bddist1}) can be violated, consider a complex $K_2$ with $n+1$ $0$-cells, called $c_0,c_1,c_2,c_3,...,c_n$ and a complex $K_1$ with $2$ $0$-cells, called $i,j$.  We pick a map $f$ that is the inverse of the map considered above.

\begin{figure}
\includegraphics[width=3in]{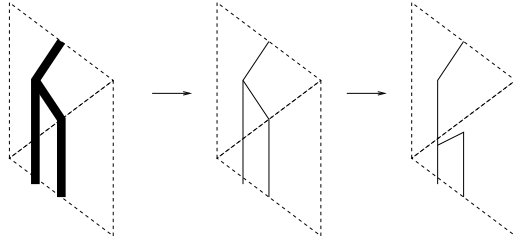}
\caption{Left image shows part of $K_2$, with dashed lines indicating $2$-cells.  Thick line is inverse image of some bad $0$-cell.  We deform the map until it is as shown in the middle image, with the inverse image of the bad $0$-cell consisting of lines with branching points; in this case there is one branching point.  Finally, we deform the map to move the branching point.  In the given case shown in this figure, at a future step we would move the branching point again through the bottom of the lower $2$-cell.  On the other hand, if we had instead drawn the lines so that one line left from the right of the bottom $2$-cell, then we would find that after the first step the bad $0$-cell would now be anchored in the bottom $2$-cell.}
\label{deform}
\end{figure}

We first show:
\begin{lemma}
Suppose $K_2$ is $1$-localizable with range $R$.  We can assume that the map $f$ from $K_2$ to $K_1$ is good at the cost of increasing $R$ by at most
a constant of order unity.
\begin{proof} 
To see that we can always assume that $f$ is good at the cost of increasing $R$ by at most $1$, first if the image of a $0$-cell in $K_2$ is not a $0$-cell, then it is a point somewhere in a $1$-cell.  So, we can simply split that $1$-cell into two $1$-cells, joined by a $0$-cell which can then be taken to be the image of the $0$-cell in $K_2$.  This does not increase the diameter of the preimage of any point in the $1$-complex, so $R$ is unchanged by this step.  

We can assume that the inverse image of every point in $K_1$ is path connected, as if the inverse image of $x$ is the union of two components $X_1,X_2$, with no path connecting one to the other, then we can define a new map.  We replace the point $x$ with two points, $x_1$ and $x_2$, and map $X_1$ to $x_1$ and $X_2$ to $x_2$, and we do similarly for a small neighborhood about $x$.

We now consider those $0$-cells in $K_1$ which are not the image of a $0$-cell in $K_2$.  We call these the ``bad" $0$-cells.
If such a bad $0$-cell in $K_1$ has degree $2$ (that is, it is attached to two $1$-cells), then it can be removed as follows: that $0$-cell is attached to two different $1$-cells, $e_1,e_2$.  We define a new complex $K'_1$, replacing those two $1$-cells with a single $1$-cell $e$, mapping $e_1$ to one half of $e$ and $e_2$ to the other half of $e$, and we map the $0$-cell to a point in the interior of $e$.
Similarly a bad $0$-cell in $K_1$ which has degree $1$ can be removed by mapping that $0$-cell and the $1$-cell attached to it onto the $0$-cell attached to the other end of the $1$-cell attached to it.  One may verify that this does not increase $R$.

Once we have removed all those bad $0$-cells with degree $1$ or $2$, consider the bad $0$-cells with degree $3$ or more.  Some such cells are as shown in Fig.~\ref{good}.  The inverse image of the $0$-cell in $K_1$ in the middle of that figure includes all $3$ $1$-cells attached to that $2$-cell in $K_2$.
In such a case, whenever the inverse image contains a path connected set in a given $2$-cell connecting all three $1$-cells in that $2$-cell, we say that the inverse image of that $0$-cell is anchored in the given $2$-cell.  Note that at most a single $0$-cell can be anchored in any given $2$-cell (there is no way to draw two different path connected sets in a given $2$-cell, both sets intersecting all three of the $1$-cells attached to the given $2$-cell).
Now, if every bad $0$-cell is anchored in some $2$-cell then we are done: since at most a single $0$-cell in $K_1$ is anchored in any given $2$-cell in $K_2$, we can pick any point in the intersection of the inverse image of that $0$-cell with that $2$-cell and call that point the center of the $2$-cell.

Now, suppose a bad $0$-cell $a$ is not anchored in any $2$-cell.  We now modify the map $f$ by deforming it.  We first deform the map so that the inverse image of every bad $0$-cell consists of lines, each being piecewise linear, with branch points allowed, as shown in Fig.~\ref{deform}.  For example, consider Fig.~\ref{good} (note that in this case the bad $0$-cell is anchored, so this deformation is not necessary, but we will describe it for this figure anyway).  In this case we modify the function so that the inverse image of the $0$-cell consists of the union of three lines, one for each $1$-cell, each line running from some point in the midpoint of the given $1$-cell in $K_2$ to some point in the middle of the $2$-cell; this point in the middle is a branch point where the three lines meet.

Note that in this process of deformation, we do not need to have ``end points" of the lines within a $2$-cell.  For example, if the intersection of the pre-image of a given bad $0$-cell with a given $2$-cell consists of a single line leaving just on one $1$-cell, with the line ending somewhere inside the $2$-cell, we can deform the map to remove that line, mapping the points that were on that line to points in one of the $1$-cells of $K_1$.

Further, we deform the function so that all branch points have degree $3$ (a branch point of degree $4$ or higher can be broken into multiple branch points of degree $3$).

Now, each bad $0$-cell has degree at least $3$, so there must be some branch point in its inverse image, and indeed there must be some branch point such that at least $3$ lines from that branch point leave the $2$-cell. For a given bad $0$-cell $a$, suppose that a branch point is contained in a $2$-cell $p$.  Suppose for simplicity that the branch point has degree $3$ and that there are no other branch points in that cell (other cases are similar).
However, since the given $0$-cell $a$ is not anchored in $p$, in fact what we must find is at least two of the lines leaving that branch point must exit that $2$-cell on the same $1$-cell on the boundary.

We now further deform the map $f$ by moving the branch point from the $2$-cell to the neighboring $2$-cell, as shown in Fig.~\ref{deform}.
We continue this process until it terminates.  We claim that this process does terminate and that this process increases $R$ by a constant of order unity.
To prove the second claim, note that the set of $2$-cells in the inverse image of every $0$-cell does not change, so the diameter cannot increase by more than twice the diameter of a $2$-cell.  To prove the first claim, one can introduce a weight function, which counts the sum over $2$-cells $p$ in $K_2$ and over $0$-cells $a$ in $K_1$ of the number of lines (counting lines joined by a branch point as separate lines) in the intersection of $p$ with the inverse image of $a$, minus the number of branch points in this intersection, and note that this weight function decreases throughout the process.
\end{proof}
\end{lemma}

If one desires, it is possible to further deform the map to get rid of those $0$-cells in $K_1$ which are not the image of a $0$-cell in $K_2$ as shown by the map from the middle to the right of Fig.~\ref{good}.  However, we are not concerned with this here.

We now show that
\begin{lemma}
Suppose $f$ is good.  
Define $l_{max}$ to be a bound on the diameter of the image of any $1$-cell in $K_2$.  Then $l_{max}$ is bounded by a function of $d,R$.
Let $D_1$ be the maximum number of $1$-cells attached to any $0$-cell in $K_1$.  Then,$D_1$ is bounded by a quantity which is a function of $d,R$.
Finally, the pre-image of every $1$-cell in $K_1$ has a diameter bounded by a function of $d,R$.
Also, 
\begin{proof}
We first show the bound $l_{max}$. The image of the $1$-cell is some path in $K_1$.  Note that given a path of length longer than $l_{max}$, it must traverse at least $l_{max}+1$ $0$-cells in $K_1$.  However, the pre-image of each such cell must be within distance $R+1$ of the start of the path (given that the length of the $1$-cell in $K_2$ is at most $1$).  Further the pre-image of each such cell must be either a $0$-cell or the center of a $2$-cell.  Hence, since the number of $0$-cells and $2$-cells within distance $R+1$ of any point is bounded by a function of $d,R$, the length $l_{max}$ is bounded.
This bound on $l_{max}$ implies Eq.~(\ref{bddist2})

The bound on $D_1$ is proven in roughly the same way as the bound $l_{max}$ on the length of the path.  Call the given $0$-cell $i$.  Since every $1$-cell has a pre-image with bounded diameter, all of the $0$-cells neighboring $i$ (that is, all the $0$-cells attached to a $1$-cell attached to $i$) have a point in their pre-image within bounded distance of the pre-image of $i$.  Since all of those pre-images of $0$-cells have bounded diameter, every point in the pre-image of those neighboring $0$-cells is within bounded distance of the pre-image of $i$.  Since each such $0$-cell has a $0$-cell or a point in the center of a $2$-cell in its pre-image and there are only a bounded number of $0$-cells and $2$-cells within any given distance of the pre-image of $i$, the bound on the number of such $0$-cells follows.

Next  we show that for a good $f$, the pre-image of every $1$-cell has a diameter bounded by a function of $d,R$.  To see this, consider some $1$-cell $e$ connecting $0$-cell $i$ to $0$-cell $j$.  The pre-image of the interior of the given $1$-cell $e$ cannot contain any $0$-cells, so the pre-image must contain a path starting and ending at $0$-cells but otherwise avoiding $0$-cells (here we use the fact that the inverse image of every point is path connected to show that this path exists).  This path $P$ can be deformed to a path $P'$ in the $1$-skeleton of $K_2$ and the image of the deformed path $P'$, by continuity, can be deformed to the path connecting $0$-cell $i$ to $0$-cell $j$ that just follows the given $1$-cell $e$.  Hence, there are two $0$-cells in $P'$ that are neighbors such that the image of one $0$-cell is $i$ and the image of the other $0$-cell is $j$; thus, since $i$ and $j$ both have bounded diameter of their pre-images, the distance between the pre-image of $i$ and $j$ must be bounded.
\end{proof}
\end{lemma}
From the above lemma, Eqs.~(\ref{bddist1},\ref{bddist2}) follow; that is, our bound on the diameter of the image or inverse image of a $1$-cell implies similar bounds on the diameter of the image of any set.  Consider, for example, any set $S_1$ in $K_1$.  We can find another set $S'_1$, with diameter at most $1$ later, such that $S_1 \subset S'_1$ and such that $S'_1$ is a union of $1$-cells.  We then use the bound on the diameter of the inverse image of each $1$-cell in $S'_1$ to bound the diameter of the inverse image of $S'_1$.

The continuity of the map $f$ is important in one other way.  
For each term $h_Z$ in the Hamiltonian, let $C(Z)$ be the subcomplex of $K_2$ containing all the $0$-cells corresponding to sites in  $Z$, all the $1$-cells
both of whose faces are $0$-cells corresponding to sites in $Z$, and all the $2$-cells attached to $1$-cells whose faces correspond to a triple of sites $i,j,k$ all in $Z$.
Every path in $C(Z)$ is contractible, and so by continuity every path in $C(Z)$ maps to a contractible path in
$f(C(Z))$.  This does not necessarily mean that $f(C(Z))$ itself is contractible.  Let us give an example; we construct this example in two steps.  Suppose one has four sites $i,j,k,l$ in $Z$.  One could map this to a $K_1$ with four $0$-cells all arranged on a line; call the corresponding $0$-cells $a,b,c,d$, with $a,b,c,d$ being the image of $i,j,k,l$ respectively and $1$-cells attached to $a$ and $b$, and to $b$ and $c$, and to $c$ and $d$.  Such a map is possible, with the $1$-cell from $i$ to $l$ being mapped to a path from $a$ to $d$.  In this case $f(C(Z))$ is contractible.  However, consider applying another map to $f(C(Z))$, mapping it to a $1$-complex with only three $0$-cells, labelled $x,y,z$ and with three $1$-cells with one $1$-cell attached to every pair of $0$-cells.  Let $a$ and $d$ be mapped to $x$; let $b$ be mapped to $y$; and let $c$ be mapped to $z$.  Map the $1$-cell attached to $c$ and $d$ to the $1$-cell attached to $x$ and $z$ and map the other $1$-cells in the natural way.  This complex now is not contractible, but still every path
in $C(Z)$ maps to a contractible path.
Also note that that given any two $C(Z)$ and $C(Z')$, their union is contractible, so any path in their union maps to a contractible path.

\section{Solving Commuting Hamiltonians on $1$-Localizable Complexes}
\label{solvesouse}
\subsection{Reducing to Commuting Projector Hamiltonians}
\label{reduce}
In this section we show that, given a locally commuting Hamiltonian with an interaction complex that is $1$-localizable, there is a
trivial state that is a ground state of the Hamiltonian, under either of two assumptions as discussed below.  First, we note that, without loss of generality, we can assume that the Hamiltonian is a locally commuting projector Hamiltonian and that any ground
state of the Hamiltonian minimizes every term $h_Z$ separately.  To see this, note that since the Hamiltonian $H$ commutes with all the
projectors $h_Z$ which all commute with each other, we can find a ground state $\psi$ that is an eigenstate of every term $h_Z$, with corresponding eigenvalue $\lambda_Z$.  Then, define for each set $Z$ a projector $P_Z$ which projects onto the eigenstates of $h_Z$ with eigenvalue $\lambda_Z$.
We define a new Hamiltonian $H'$ to be the sum of $I-P_Z$ over all $Z$.  Then, $\psi$ is a ground state of this new Hamiltonian $H'$ with zero energy
and any ground state $\phi$ of this Hamiltonian $H'$ is also a ground state of $H$ since it has the same expectation value for energy.
Of course, defining this Hamiltonian $H'$ does require knowledge of the $\lambda_Z$; however, in this section we are not concerned with the question of how to determine the $\lambda_Z$ but rather with whether or not the Hamiltonian $H$ has a trivial ground state and if we can show that $H'$ has a trivial ground state it will follow that $H$ has a trivial ground state.  Thus, if for some interaction complex every local commuting projector Hamiltonian has a trivial ground state, then for that interaction complex every local commuting Hamiltonian has a ground state.
We begin in subsection \ref{2site} with a review of the method of Ref.~\onlinecite{bv}, applicable to the case where the Hamiltonian is a sum of two-site and one-site interaction terms before considering the more general case in subsection \ref{gs1l}.

\subsection{Two-Site Commuting Hamiltonians}
\label{2site}
Suppose we have a Hamiltonian which is a sum of two-site and one-site interaction terms.  We write this as
\be
\label{twobody}
H=\sum_{<i,j>} H_{i,j}+\sum_{i} H_{i,i},
\ee
where $H_{i,j}$ acts only on sites $i,j$ and $H_{i,i}$ acts only on site $i$.  Suppose all the various terms $H_{i,j}$ and $H_i$ commute with
each other.  Consider any site $i$.
Decompose $H_{i,j}$  as a sum of product operators $H_{i,j}=\sum_\gamma O_i^{ij}(\gamma) O_j^{ij}(\gamma)$, where the operators $O_i^{ij}(\gamma),O_j^{ij}(\gamma)$ 
are supported on $i,j$ respectively and the operators $O_j^{ij}(\gamma)$ are chosen from an orthonormal basis.
 Then, $[O_i^{ij}(\delta),O_i^{ik}(\gamma)]=0$ for 
$j\neq k$, for all $\delta,\gamma$.  
Let ${\cal A}^{ij}$ be the algebra generated by the set of $O_i^{ij}(\gamma)$ for given $j$.  The algebras ${\cal A}^{ij}, {\cal A}^{ik}$ commute for $j \neq k$.

This concept of interaction algebra originates in Ref.~\onlinecite{intalg}.  Generally, given any operator $O$ and any set $X$ we say that the interaction algebra of $O$ on $X$ is the algebra supported on $X$ generated by ${\rm tr}_{\overline X}(O Q_{\overline X})$ where the trace is over all sites in the complement of $X$ and $Q_{\overline X}$ is any operator supported on the complement of $X$.

Let ${\cal H}_i$ denote the Hilbert space on site $i$.  
Then, it is a fact from $C^*$-algebra that
we can decompose ${\cal H}_i$ into a direct sum of Hilbert spaces ${\cal H}_i^{\alpha(i)}$,
\be
\label{decompsum}
{\cal H}_i=\bigoplus_{\alpha(i)} {\cal H}_i^{\alpha(i)},
\ee
 and then further decompose each such Hilbert space ${\cal H}_i^{\alpha(i)}$ into a tensor product of spaces ${\cal H}_{i\rightarrow j}^{\alpha(i)}$ (where the product ranges over $j$ that neighbor $i$) tensor producted with space ${\cal H}_{i, i}^{\alpha(i)}$ so that
\begin{eqnarray}
\label{decomp1}
{\cal H}_i& = &\bigoplus_{\alpha(i)} {\cal H}_i^{\alpha(i)}
= \bigoplus_{\alpha(i)} \Bigl( {\cal H}_{i,i}^{\alpha(i)} \otimes \bigotimes_{<j,i>} {\cal H}_{i\rightarrow j}^{\alpha(i)} \Bigr),
\end{eqnarray}
where the product is over $j$ that neighbor $i$,
such that each operator $H_{i,j}$ can be decomposed as
\be
\label{decomp2}
H_{i,j}=\sum_{\alpha(i),\beta(j)} P_i^{\alpha(i)} P_j^{\beta(j)} H_{i,j}^{\alpha(i),\beta(j)},
\ee
where $P_i^{\alpha(i)}$ is the operator on ${\cal H}_i$ which projects onto ${\cal H}_i^{\alpha(i)}$ and $H_{i,j}^{\alpha(i),\beta(j)}$ acts
on the subspace of ${\cal H}_i^{\alpha(i)} \otimes {\cal H}_j^{\beta(j)}$ given by ${\cal H}_{i\rightarrow j}^{\alpha(i)} \otimes {\cal H}_{j \rightarrow i}^{\beta(i)}$ and such that
$H_{i,i}$ can be decomposed as
\be
\sum_{\alpha(i)} P_i^{\alpha(i)} H_{i,i}^{\alpha(i)}
\ee
where $H_{i,i}^{\alpha(i)}$ acts only on ${\cal H}_{i,i}^{\alpha(i)}$.

To give an example, 
suppose that there is only one possible choice of index $\alpha(i)$, 
so that Eq.~(\ref{decompsum}) has only one term on the right-hand side.  Then
${\cal H}_i$ decomposes into a tensor product of Hilbert spaces, and each ${\cal A}^{ij}$ acts on a different space.  This would include a case, for example, in which the Hilbert space on site $i$ had dimension $4$, and decomposed into the tensor product of two spin-$1/2$ degrees of freedom, one such spin-$1/2$ degree of freedom interacting with some site $j$ and the other one interacting with some site $k$.

To give the simplest example with multiple terms in Eq.~(\ref{decompsum}),
consider a single spin-$1/2$ degree of freedom on each site, with an Ising Hamiltonian with all terms involving only operators $S^z$.  
Then, there are two different terms in the sum of Eq.~(\ref{decompsum}), corresponding to the spin up and spin down states and each Hilbert space ${\cal H}_i^{\alpha(i)}$ is $1$-dimensional.

To understand the operators $P_i^{\alpha(i)}$, it is important to understand the concept of ``central elements" of an algebra.  These are elements of the algebra that commute with every other element of the algebra.  All algebras contain the identity operator as a central element, but sometimes other, nontrivial, central elements may be present as well.  The decomposition Eq.~(\ref{decomp1}) can be effected by choosing the $P_i^{\alpha(i)}$ to be central elements in the algebra generated by all the interaction algebras on site $i$; that is, consider the various interaction algebras ${\cal A}^{ij}$, one such algebra for each $j$ which is a neighbor of $i$, and take the algebra generated by those algebras and then take the central elements of that algebra.  If there are no central elements, then we have only one choice of index $\alpha(i)$ and we can directly decompose ${\cal H}_i$ into a tensor product so that each ${\cal A}^{ij}$ acts on only one factor in the tensor product.  The case of an Ising Hamiltonian above has the projectors onto the spin up or spin down states being nontrivial central elements.

Given a Hamiltonian of form (\ref{twobody}), the operators $P_i^{\alpha}$ commute with each other and commute with the Hamiltonian for any $i$ and any $\alpha$.  Thus,
we can assume that the ground state is an eigenstate of all of these operators.  A state $\psi$ that is an eigenstate of all of these operators has the following form: for each site $i$, there is some $\alpha(i)$ such that $P_i^{\alpha(i)} \psi=\psi$ and $P_i^\beta \psi=0$ for all $\beta\neq \alpha(i)$.   
Given the decomposition (\ref{decomp2}), the ground state on such a graph has very simple entanglement properties: 
the ground state is a product of states $\psi_{i,j}$, where $\psi_{i,j}$ is in
the space ${\cal H}_{i \rightarrow j}^{\alpha(i)}\otimes {\cal H}_{j \rightarrow i}^{\beta(j)}$, and states $\psi_{i}$ in ${\cal H}_{i,i}^{\alpha(i)}$.
This state can be created by acting on a product state with a unitary quantum circuit as follows.  On each round of the quantum circuit, each unitary gate in that round acts only on some given nearest neighbor pair $i,j$ (and hence the diameter of the sets it acts on is equal to $1$).  We must have enough rounds that each edge in the graph has some corresponding unitary and such that in any given round there are no two gates acting on the same site.  To do this, define several sets of edges, called $Y_1,...,Y_k$, such that each edge appears in exactly one such set and such that no two edges in a given set connect to the same site.  We can find such sets using a $k$ that is bounded by a function of the degree $d$ of the graph (to see this, proceed in a greedy fashion, constructing the set $Y_a$ by greedily adding edges that are not in any previous set and that do not share a site with any other edge already added to $Y_a$, increasing the size of $Y_a$ until this greedy procedure terminates; then if an edge is not in $Y_a$, some other edge that shares a site with that edge must be in $Y_a$; since there are only at most $2(d-1)$ edges that share a site with any given edge and since each edge is only in one of these sets $Y_1,...,Y_k$, after $2d-1$ rounds all edges must be chosen).  One way to regard the problem of finding these sets is as a graph coloring problem: we must color each edge with one of $k$ colors so that no two edges sharing a site share the same color.
Then, in the first round we choose unitaries acting on pairs of sites such that the edges connecting those sites are in $Y_1$; in the second round we act on those pairs such that the edges are in $Y_2$; and so on.
This gives a quantum circuit with $k$ rounds so if the graph has bounded degree this is a trivial state.

\subsection{Ground States on $1$-Localizable Complexes}
\label{gs1l}
We now show how to construct the ground state given that the interaction complex is $1$-localizable under either a technical assumption on the Hamiltonian, or an assumption that the girth of the complex $K_1$ is sufficiently large.
We first consider the high girth case:
\begin{theorem}
Consider a local commuting Hamiltonian with interaction complex $K_2$ with degree $d$ which is $1$-localizable with range $R$.  
Let  $f$ be the function used to show that $K_2$ is $1$-localizable in definition (\ref{defn1}), and let $K_1$ be the image of $K_2$ under $f$.  Let us assume $f$ is good.  Assume the girth of $K_1$ is greater than $2 l_{max}$, where $l_{max}$ is the maximum diameter of the image of any $1$-cell of $K_2$.
(Note that this  includes the graphs of section \ref{countersubsec} which cannot be coarse-grained into a triangle-free graph.)
Then $H$ has a trivial ground state constructed by a quantum
circuit with depth and range both bounded by functions of $d,R$, using ancillas with the dimension of the ancilla on a given site bounded by a function of $d,R$.
\begin{proof}
Consider any Hamiltonian with the following form:
\be
\label{formabove}
H'=\sum_{x\in J} h'(x),
\ee
where the sum is over some set $J$ of points $x$ (each point $x$ is not necessarily a $0$-cell) in $K_1$ and $h'(x)$ is supported on the set of sites $i$ such that the $0$-cell corresponding to $i$ in $K_1$ is within distance $l_{max}/2$ of $x$.   
We make the following two claims.  First, because the girth of $K_1$ is sufficiently large, the interaction $2$-complex corresponding to $H'$ can be continuously mapped to $K_1$ for any choice of $H'$.  The proof of this claim is essentially the same as our previous proof below definition \ref{defn1} that the graphs $G^R$ constructed their are $1$-localizable.  Given a Hamiltonian of the form $H'$, the interaction complex has free edges.  For example, the longest edge (longest as measured by the distance on $K_1$ between its endpoints) is free (that is, this edge is attached to no other edges).  One can then map this edge to remove a $2$-cell; one can repeat this process until one has just a $1$-complex left.
Secondly, we claim that
if the image of any $1$-cell in $K_2$ has diameter at most $l_{max}$ then $H$ can be written in the form (\ref{formabove}), since each interaction term in $H$ is supported on a set $Z$ whose image in $K_1$ has diameter at most $l_{max}$.  Note that since $K_1$ is high girth, the set of points within distance $l_{max}$ of any point is a tree, so that a set with diameter at most $l_{max}$ contains only points within distance $l_{max}/2$ of some given point.

Let $H_0=H$, $f(0)=f$, and $K_1(0)=K_1$.
We  give an iterative construction of a trivial ground state of $H_0$: we will start with Hamiltonian $H_0$, and then define a new Hamiltonian, $H'_0$, such that any ground state of $H'_0$ will be a ground state of $H_0$ (we will do this by taking $H'_0$ to equal $H_0$ plus an additional projector supported on some set $Z$ with diameter at most $l_{max}$), and then we will show how to define another Hamiltonian, $H_1$, which acts both on the original degrees of freedom as well as certain additional ancilla degrees of freedom.  The interaction complex corresponding to $H_1$ is still $1$-localizable but is mapped by a {\it different} function $f(1)$ to a {\it different} $1$-complex, $K_1(1)$ such that there is a unitary transformation $U(1)$ turning any ground state of $H_1$ into a state on the original degrees of freedom plus the ancilla degrees of freedom such that its state on the original degrees of freedom is a ground
 state of $H_0'$.  The $1$-complex $K_1(1)$ will (in some sense defined below) be ``simpler" than $K_1$; we will then repeat this procedure applied to Hamiltonian $H_1$, and continue iteratively, until we have some Hamiltonian defined on a sufficiently simple $1$-complex that the ground state is a product state.  Then, we will apply the product of unitaries $U(1) U(2) ...$ to this product state to construct a trivial ground state of $H$.  We will then show that this product of unitaries can be written as a local quantum circuit with ancillas.

To construct $H'_0$ from $H_0$, first write $H_0=\sum_{x \in J} h_0(x)$ as in the form Eq.~(\ref{formabove}).  Then, consider any $0$-cell, $i$, in $K_1(0)$.
We refer to this as ``choosing" the $0$-cell $i$ on the given step of the iterative construction and we describe below how to appropriately choose the $0$-cells on each step.
Let $Z$ be the set of sites whose corresponding $0$-cells in $K_1(0)$ are within distance $l_{max}/2$ of $i$.
The number of $1$-cells attached to $i$ is bounded by some quantity $D_1$, where $D_1$ is bounded by a function of $d,R$. 

Label the $1$-cells attached to $i$ by $i_1,i_2,...,i_n$, where $n\leq D_1$.  Define
$H_0(i_j)$ to be the sum of $h_0(y)$ over all $y\in J$ with $y \neq i$ such that ${\rm dist}(y,i) \leq l_{max}/2$ and such that 
the shortest path from $i$ to $y$ includes some point in the $1$-cell $i_j$ other than the point $i$ (i.e., $y$ is in the $1$-cell $i_j$ or the shortest path from $i$ to $y$ contains the $1$-cell $i_j$).  Because of the lower bound on girth, there is a unique shortest path for these points $y$ with ${\rm dist}(y,i)\leq l_{max}/2$.
Then, $\sum_j H_0(i_j)$ contains all interaction terms on $H$ with support on $Z$ (other than $h_0(i)$ if $i \in J$).  (See also the appendix and discussion of shields for additional discussion of these terms $H_0(i_j)$ which may be useful).
In Fig.~\ref{tree}, if the point $i$ is the solid circle on the left, then there are three distinct terms $H_0(i_1),H_0(i_2),H_0(i_3)$ corresponding to the three $1$-cells attached to $i$.

We claim that the interaction algebras of the different $H_0(i_j)$ on $Z$ commute with each other, and also commute with $h_0(i)$, if $i \in J$.  To show this, it suffices to consider any $h_0(y)$ which appears in the sum defining $H_0(i_j)$ and some other $h_0(x)$ which appears in the sum defining $H_0(i_k)$ for $j \neq k$ and show that the interaction algebras of $h_0(y)$ and $h_0(x)$ on $Z$ commute with each other.  
Let $I(y,x)$ be the intersection of the support of $h_0(y)$ and the support of $h_0(x)$.  Note that since $h_0(y)$ commutes with $h_0(x)$, the interaction algebra of $h_0(y)$ on $I(y,x)$ commute with the interaction algebra of $h_0(x)$ on $I(y,x)$.
Hence, if the interaction algebras on $Z$ fail to commute then $I(y,x)$ has non-vanishing intersection with the complement of $Z$; however, also if the interaction algebras on $Z$ fail to commute then $I(y,x)$ has non-vanishing intersection with $Z$
(if not, the interaction algebras would trivially commute as the support of one interaction algebra would be disjoint from the support of the other interaction algebra).  However, if $I(y,x)$ has non-vanishing intersection with both $Z$ and the complement of $Z$, this contradicts the assumption on the lower bound on girth (alternately, this contradicts the assumption that the interaction complex is $1$-localizable, which is what we will use for arbitrary complexes later).

Given that these interaction algebras commute, by the ideas discussed in the previous subsection we can decompose the Hilbert space ${\cal H}_Z$ on $Z$ into a direct sum of Hilbert spaces ${\cal H}_Z^\alpha$, such that each ${\cal H}_Z^\alpha$ decomposes into a product of subspaces ${\cal H}_{Z\rightarrow i_j}^\alpha$ and subspace ${\cal H}^\alpha_{Z,Z}$ such that the interaction algebra $H_0(i_j)$ on $Z$ acts only on spaces ${\cal H}_{Z \rightarrow i_j}^{\alpha}$ and $h_0(i)$ acts only on ${\cal H}^\alpha_{Z,Z}$.  There is some subspace ${\cal H}_Z^\alpha$ for some given $\alpha$ and some projector $P_Z^\alpha$ onto that subspace such that a ground state $\psi$ of $H_0$ obeys $P_Z^\alpha \psi=\psi$.  Define $H'=H+(1-P_Z^\alpha)$.  Note that the projector $1-P_Z^\alpha$ can be written as a term supported on a set of sites within distance $l_{max}/2$ of a given site so the Hamiltonian $H'$ still is in the form of Eq.~(\ref{formabove}).

To define $H_1$, we add additional ``ancilla" degrees of freedom: for each site in $Z$, we add up to $D_1-1$ additional ancilla degrees of freedom.  In fact, there
is no particular reason to refer to one of those degrees of freedom as ``real" and the others as ``ancillas": each site in $Z$ will now have a total of up to $D_1$ copies, labelled $1,2,...,$.  To define $H_1$, all interaction terms in $H_0$ without support on $Z$ appear in $H_1$ also.  The term $1-P_Z^\alpha$ in $H'_0$ is replaced in $H_1$ by a sum of up to $D_1$ terms, one term acting on each copy of $Z$.  That is, we replace it with $\sum_{j} (1-P_{Z(j)}^\alpha)$, where $P_{Z(j)}^\alpha$ is supported on the $j$-th copy of the sites in $Z$ and the sum is over copies.  Similarly, any interaction term in $H_0$ which is supported on $Z$ (i.e., it has no support outside $Z$) is replaced by a sum of that term acting on each copy.  Finally, any interaction term in $H_0$ which has support on $Z$ but which is not given supported on $Z$ must be $h_0(y)$ for some $y \neq i$ such that the shortest path from $i$ to $y$ includes some point in the $1$-cell $i_j$ other than $i$; such interaction terms $h_0(y)$ are replaced by that interaction term acting on the $i_j$-th copy of the sites in $Z$ (i.e., take that term $h_0(y)$ and replace any operator acting on a site in $Z$ by the corresponding operator acting on the site in the $i_j$-th copy of that site).
This defines $H_1$.  We define the new complex $K_1(1)$ by a procedure as exemplified in Fig.~\ref{tree}: in an abuse of language, we say that a $0$-cell is in $Z$ if it corresponds to a site in $Z$ and then the $0$-cells in $Z$ are replaced by up to $D_1-1$ copies.  If there is an edge in $K_1(0)$ between two $0$-cells $i$ and $j$ in $Z$, then we have an edge in $K_1(1)$ between the corresponding $0$-cells if they are in the same copy.  If there is an edge in $K_1(0)$ between a $0$-cell $i$ in $Z$ and a $0$-cell $j$ not in $Z$, then we have an edge between the corresponding $0$-cells in $K_1(1)$ depending upon whether there is an interaction term in $H_1(1)$ acting on the corresponding sites or not (which depends upon the copy and upon $i$ and $j$).
Again, $H_1$ can be written as a Hamiltonian in the form Eq.~(\ref{formabove}) except now we use the complex $K_1(1)$ instead of $K_1(0)$: the points $x$ are in $K_1(1)$ and the support of operators $h'(x)$ is on sites whose corresponding $0$-cell is within distance $l_{max}/2$ of $x$ in $K_1(1)$.

Note that there is a unitary $U(1)$ that maps any ground state of $H_1$ into a state on the original and ancilla degrees of freedom such that that state is a ground state of $H'_0$ on the ancilla degrees of freedom.  To construct this unitary, simply swap, for each $j>1$, the state on ${\cal H}_{Z(i_j) \rightarrow i_j}$ with ${\cal H}_{Z(i_1) \rightarrow i_j}$.  Then, let $i_1$ be the real degree of freedom and the other $i_2,i_3,...$ denote the ancilla degrees of freedom.
This unitary has bounded range.

We repeat this procedure until we arrive at a Hamiltonian $H_n$ such that $K_1(n)$ is a tree graph.  Then, we can construct a trivial ground state on $K_1(n)$ by coarse-graining $H_n$ to a Hamiltonian such that all terms act on at most two sites.  We now explain how to choose sites $i$ at each step such that the procedure will terminate and such that $U(1) U(2) ... U(n)$ will be a quantum circuit with a bound on the depth.

We can guarantee that this procedure will terminate, since if $K_1(l)$ is not a tree for some $l$, then we can find an $i$ such that some cycle contains $1$-cells attached to $i$.  Then, for this $i$, the first Betti number of $K_1(l+1)$ will be less than that of $K_1(l)$.
Note that after the first step of this procedure, some of the sites will have several copies, corresponding to the ancilla degrees of freedom that we added.
Suppose a given site $k$ is in the set $Z$ on some step.  If $k=i$, then we will never choose any of the (at most) $D_1$ copies of $k$ on any future step.  If $k$ is in $Z$ but $k\neq i$, then if the shortest path from $i$ to $k$ goes contains the $1$-cell $i_a$ then we only will choose the $a$-th copy of $k$ as that is the only copy such that choosing that copy will reduce the first Betti number.
See Fig.~\ref{tree}.

\begin{figure}
\includegraphics[width=3.1in]{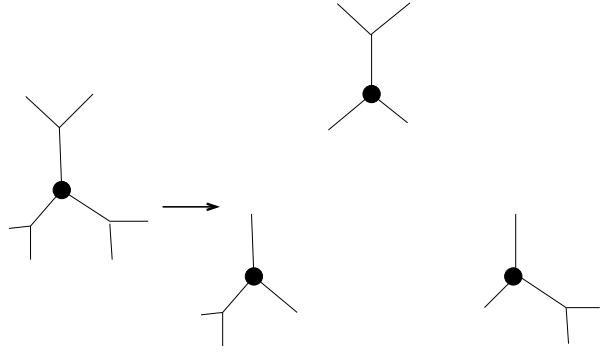}
\caption{Illustration of how the graph is changed.  On the left, we illustrate part of the graph, with the lines indicating edges.  Vertices are {\it not} shown to avoid cluttering the graph, but the vertex $i$ is shown as a solid circle.  The set $Z$ contains $i$ as well as its first neighbors.  To the right of the arrow, we show how the tree is transformed, with each site in $Z$ having three copies.}
\label{tree}
\end{figure}

So, when we say that we ``choose" a site $i$ on some given round of this procedure, that site $i$ may possibly be some copy of a site $i_0$ on a previous round.  The number of copies that $i_0$ has may change from round to round.  However, at any given round there is only one copy of such a site $i_0$ that we can choose to reduce the first Betti number.  So, when we describe how to choose sites, we will simply specify which site in the original graph that we choose, and then we assume that we pick the appropriate site in the graph after the given number of steps.

So we can now describe how to choose sites.
Define several sets of sites $Y_1,Y_2,...,Y_k$, such that all sites in $Y_1$ have distance greater than $l_{max}$ from each other.  Because of the bound on the degree of the graph, it is possible to find such sets of sites such that every site appears in exactly one such set using a $k$ that is bounded by a function of $d,R$.  So, we order the sites $i$ that we choose as follows: the first steps choose all the sites in $Y_1$, choosing one such site on each step; then we choose all those in $Y_2$, and so on.  Then, all the unitaries result from sites in $Y_1$ can be considered to be the first round of the quantum circuit (since the supports of those unitaries do not overlap), all those unitaries resulting from sites in $Y_2$ can be considered to be the second round, and so on, giving a quantum circuit with $k$ rounds.
Thus, we have succeeded in showing that the ground state is trivial in this case.
\end{proof}
\end{theorem}

This construction used ancillas.  One might wonder if they are necessary.  For example, certain states in two-dimensional quantum systems such as the ground state of a Chern insulator cannot be approximated by trivial states but have the property that a state which is a product of
the ground state and its complex conjugate (to cancel certain K-theory obstructions) can be approximated by a trivial state.
In the present case, however, the ancillas serve primarily a bookkeeping purpose.  In the definition of $H_1$, note that the interaction algebra on $H_1$ on the $k$-th copy of $Z$  on the subspace ${\cal H}_{Z \rightarrow i_j}^\alpha$, for $k\neq j$, is either trivial or generated by commuting projectors (that is, all terms in this algebra are central).  Thus, we could choose to add to $H_1$ additional interaction terms for each $j,k$ for $j\neq k$, of the form $1-P_{Z,j,k}$, where $P_{Z,j,k}$ acts on the subspace ${\cal H}_{Z \rightarrow i_j}^\alpha$ and is a rank-$1$ projector.  These terms commute with all other terms in the Hamiltonian.  They ensure that the ground state is in the range of $P_{Z,j,k}$ for each $j,k$.
By doing this, we ensure that the final state of the ancillas is ``trivial" also, in that if
 $U(\psi_{prod}^{real}\otimes \psi_{prod}^{ancilla})=\psi_{out}^{real} \otimes \psi^{ancilla}$, then $\psi_{out}^{real}$ is the desired ground state and $\psi^{ancilla}$, the final state of the ancillas, is a trivial state without the use of further ancillas (i.e., the state $\psi^{ancilla}$ can be constructed by a unitary quantum circuit of bounded depth and range applied to a product state without using further ancillas).  This is very distinct from the case of a Chern insulator, where the ancillas are used to cancel K-theory obstructions and the final state of the ancilla cannot be constructed by a bounded depth and range circuit applied to a product state without using further ancillas.  In fact, the ancillas play primarily a ``book-keeping" role to keep track of different subspaces of the Hilbert space and can be avoided, but we do not discuss this further.

Now, suppose that $K_1$ does {\it not} have high girth.  It is still possible in this case to find trivial ground states for many Hamiltonians, under one technical assumption which holds for all stabilizer Hamiltonians.
The main difficulty that we encounter in studying these systems is the presence of central elements in the interaction algebra.  In the theorem above, we dealt with these central elements by adding terms $1-P_Z^\alpha$ that project orthogonal to the desired subspace.
We deal with these differently here.
Before giving this approach, let us define what we main by stabilizer Hamiltonians.
\begin{definition}
A locally commuting projector Hamiltonian is a stabilizer Hamiltonian if the Hilbert space on every site is a tensor product of $2$-dimensional Hilbert spaces, called qubits, and if every projector $h_Z$ in Eq.~(\ref{Hsum}) has the form
\be
h_Z=\frac{1 \pm O^{Pauli}_Z}{2},
\ee
where $O^{Pauli}_Z$ is a product of Pauli operators acting on the qubits in set $Z$.  The sign, $\pm$, can be chosen arbitrarily for each set $Z$.
\end{definition}
We remark that commonly one defines a stabilizer Hamiltonian to simply be a sum of terms $\pm O^{Pauli}_Z$; we instead define the Hamiltonian as above so that it will be a sum of commuting projectors.  Any state $\psi$ which is a zero eigenvector of $h_Z$ obeys $\mp O^{Pauli}_Z \psi=\psi$.

The approach we describe will allow us to show that
\begin{theorem}
\label{stabtriv}
Consider a local commuting stabilizer Hamiltonian with interaction complex $K_2$ with degree $d$ which is $1$-localizable with range $R$.  
Let  $f$ be the function used to show that $K_2$ is $1$-localizable in definition (\ref{defn1}), and let $K_1$ be the image of $K_2$ under $f$.  Let us assume $f$ is good.  
Then $H$ has a trivial ground state constructed by a quantum
circuit with depth and range both bounded by functions of $d,R$, using ancillas with the dimension of the ancilla on a given site bounded by a function of $d,R$.
\end{theorem}

We postpone the proof of this theorem briefly.  The reason is,
rather than just describing the approach in the special case of stabilizer Hamiltonians, we begin by describing the approach for more general Hamiltonians.
We will show that if one can construct certain Hamiltonians, Eq.~(\ref{cross1}), with properties that we give below, then the Hamiltonian has a trivial ground state.  Finally, we show that such a construction can be done for stabilizer Hamiltonians, provng theorem \ref{stabtriv}.

\begin{figure}
\includegraphics[width=1.9in]{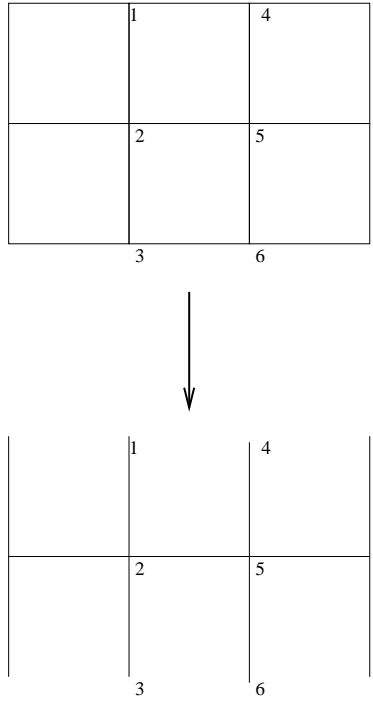}
\caption{Illustration of ``cut".  Top shows a system of $12$ sites, with the 6 sites in the center labelled $1,2,3,4,5,6$ as shown.  Each interaction term acts on $4$ sites in a square (to be consistent with our previous definition of interaction complexes, we should have shown diagonal lines going across each square, but we have left those out to avoid cluttering the image).  Bottom shows map to a $1$-complex.  We cut the system across the middle cut, to product a term $H_{LR}$ acting on sites $1,2,3,4,5,6$, and terms $H_{LO}$ and $H_{RO}$, with $L$ being sites $1,2,3$ and $R$ being sites $4,5,6$.}
\label{figcut}
\end{figure}

Assume $f$ is good.
We begin by describing how to ``cut" the complex $K_2$ by removing points in some set, this set being a pre-image of the interior of some $1$-cell in $K_1$.  
For technical reasons, we go to the cover when describing how to cut the complex, because of the possibility mentioned before that $f(C(Z))$ might not be contractible.  The goal of this ``cutting" is to find a certain decomposition of the Hamiltonian as in Eq.~(\ref{Hlro}) and as in Fig.~\ref{figcut} given later.

The universal cover of $K_1$ is a tree, $\tilde K_1$; if $K_1$ has cycles, then $\tilde K_1$ is an infinite tree.  Let $h$ be the covering map from $\tilde K_1$ to $K_1$.  Consider a point $x$ in a $1$-cell in $K_1$, $\sigma$ (we describe how to choose $x$ below) with $x$ not contained in a $0$-cell.  Let $\tilde x$ be a point in the pre-image under $h$ of $x$ (we describe how to choose $\tilde x$ below).  The point $\tilde x$ is contained in some $1$-cell, called $\tilde \sigma$.  Since $\tilde K_1$ is a tree, removing the point $\tilde x$ divides $\tilde K_1$ into two trees, which we call the ``left" and ``right" trees.  Consider each $h_Z$ such that the pre-image under $h$ of $f(C(Z))$ contains $\tilde x$.  Consider each site $i \in Z$,
and take a path in $C(Z)$ starting at $x$ and ending at the $0$-cell corresponding to $i$.  This path lifts to some path in $\tilde K_1$ starting at $\tilde x$ and ending at a pre-image of the $0$-cell corresponding to $i$.  If this pre-image of the $0$-cell can be deformed to a path entirely in the left tree, then call $i$ a ``left site for $h_Z$" and otherwise call $i$ a ``right site for $h_Z$".
We claim that this procedure is consistent: for every site $i$, if $i$ is a left site for some $h_Z$, then it will not be a right site for any other $h_{Z'}$.  This follows from the continuity: since the union of $C(Z)$ and $C(Z')$ is contractible, a closed path starting at $x$, moving to $i$ in $C(Z)$, and returning to $x$ in $C(Z')$ is a contractible path and hence maps to a closed contractible path in $K_1$ and hence a closed path in $\tilde K_1$.

We say that a site $i$ is a left site if it is a left site for some $h_Z$ and we say that $i$ is a right site if it is a right site for some $h_Z$.  Otherwise, we say that $i$ is ``other".  Some of the interaction terms $h_Z$ have the property that the set $Z$ contains both left and right sites.  For such sets $Z$, every site in $Z$ is either a left site or a right site, not an other site (as if $Z$ contains both left and right sites, then the pre-image of $Z$ intersects $\tilde x$).
Let $H_{LR}$ be the sum of all such terms $Z$.  Other interaction terms involve only left and other sites or only right and other sites; we denote the sum of the terms involving only left and other sites as $H_{LO}$ and the sum of terms involving only right and other sites as $H_{RO}$.
So,
\be
\label{Hlro}
H=H_{LR}+H_{LO}+H_{RO}.
\ee
Let $L$ and $R$ be the set of left and right sites, respectively, with ${\cal H}_L$ and ${\cal H}_R$ being the Hilbert spaces on these sites.  Note that the dimensions of these Hilbert spaces, ${\cal H}_L$ and ${\cal H}_R$, is independent of system size and depends only upon $d,R$, so that operations on them can be done efficiently.
See Fig.~\ref{figcut} for an illustration.

So, as before we can define an interaction algebra of $H_{LR}$ on the left and right sites.  This interaction algebra may have nontrivial central elements.  If it does {\it not} have nontrivial central elements, then we can proceed as follows.  Let ${\cal H}_{L\rightarrow R}$ and ${\cal H}_{R \rightarrow L}$ be subspaces of the Hilbert spaces ${\cal H}_L$ and ${\cal H}_R$ on $L$ and $R$ such that $H_{LR}$ acts on ${\cal H}_{L\rightarrow R}\otimes {\cal H}_{R \rightarrow L}$ and $H_{LO}$ and $H_{RO}$ do not act on that space.
Add additional ancilla degrees of freedom on $L$ and $R$ initialized to a given state.   Apply a unitary transformation supported on $L$ and $R$ to swap ${\cal H}_{L \rightarrow R}\otimes {\cal H}_{R \rightarrow L}$ between real and ancilla degrees of freedom and then transform that entangled state on the ancilla degrees of freedom to the given product state.  This gives a  unitary transformation that maps any ground state of $H_{LO}+H_{RO}$ tensored with the given product state on the ancillas to a ground state of $H$.
Similarly to the high girth case, we proceed iteratively: find a sequence of different points $x$, and a corresponding sequence of Hamiltonians and unitary transformations.  The idea is that the Hamiltonian $H_{LO}+H_{RO}$ has the same interaction complex as the orginal one, except it has been ``cut open" across the cut; we repeat the procedure until we have an interaction complex that can be mapped to a tree, at which point we have a trivial ground state.

The only problem that can arise in this procedure is if there are nontrivial central elements in the interaction algebras.
We now encounter a difficulty.  Decompose ${\cal H}_L$ into a direct sum of subspaces ${\cal H}_L^\alpha$ (and similarly decompose ${\cal H}_R$) so that each term in the sum decomposed into a product of two subspaces with $H_{LR}$ and $H_{LO}$ acting on different subspaces in that product.  We would like to proceed as before, swapping degrees of freedom between the real and ancilla spaces.  However, we need to ensure that these degrees of freedom are in the correct subspaces ${\cal H}_L^\alpha$ and ${\cal H}_R^\alpha$.  We can do this by adding a term to the Hamiltonian projecting on a given subspace of ${\cal H}_{L}$.  However, adding such a term $1-P^\alpha_L$ may ruin the properties of the Hamiltonian.  Adding this term may mean that the interaction complex corresponding to $H+(1-P^\alpha_L)$ is no longer $1$-localizable with range $R$ and may  also increase the degree $d$ of $K_2$.

The approach we try to resolve this is to add ancilla degrees of freedom and make the projectors act on the ancillas instead.  The particular form of the Hamiltonian below Eq.~(\ref{cross1}) is chosen so that the interactions that couple the real degrees of freedom (the real degrees of freedom are the degrees of freedom other than the ancillas; i.e., they are the original degrees of freedom) to the ancillas take a similar form as the interactions in the original Hamiltonian, so that we do not destroy the $1$-localizability properties of the interaction complex.  However, since we have imposed the projectors on the ancilla degrees of freedom, we must make sure that the couplings of the real degrees of freedom to the ancillas are sufficiently restrictive to fix the values of the real degrees of freedom as needed.  This is not always possible (as a toy example that we give below shows), but in the special case of stabilizer Hamiltonians we show that this is indeed possible.
We now explain this approach in more detail.
Add ancilla degrees of freedom on $L$ and $R$.  Let $L^r$ denote the set of real degrees on $L$, $L^a$ denote the ancilla degrees of freedom on $L$, and similarly for $R^r$ and $R^a$.  Now, try to construct a Hamiltonian of the form
\be
\label{cross1}
H_{L^rR^a}+H_{L^aR^r}+H_{L^rO}+H_{R^rO}+(1-P_{L^a})+(1-P_{R^a}),
\ee
where all terms commute with each other.
Here $H_{L^rO}$ is the same as the Hamiltonian $H_{LO}$ above, except acting on the real degrees of freedom on $L$.  $H_{R^rO}$ is defined similarly.  The term $H_{L^rR^a}$ is the same as term $H_{LR}$ except acting on the real degrees of freedom on $L$ and the ancilla degrees of freedom on $R$.  Finally, $P_{L^a}$ and $P_{R^a}$ are suitably chosen projectors on the ancilla degrees of freedom (we describe how they are chosen below).
We would like to choose the projectors $P_{L^a}$ and $P_{R^a}$ such that we can find a unitary acting on the real and ancilla degrees of freedom on $L,R$ such that any ground state of such a Hamiltonian (\ref{cross1}) is mapped to a ground state of $H$.

Note that if we choose the unitary so that 
it commutes with all elements in the interaction algebra of $H_{L^rO}$ on $L^r$ and the interaction algebra of $H_{R^rO}$ on $R^r$, including in particular the central elements, then it will map every ground state of Eq.~(\ref{cross1}) to a ground state of $H_{L^rO}+H_{R^rO}$; this still leaves us with the problem of ensuring that the unitary will also map the ground states to ground states of $H_{LR}$.

Suppose one could construct such a Hamiltonian for every point $x$ and every pre-image.  This means that if we can construct a ground state of the Hamiltonian Eq.~(\ref{cross1}), we can construct the ground state of the original Hamiltonian by applying a local unitary.  The interaction complex corresponding to Hamiltonian Eq.~(\ref{cross1}) differs from the original interacton complex in that it has been ``cut open" on the cut and possibly additional interaction terms have been added on both sides of the cut; the effect on the interaction complex is slightly different from before.
We apply an iterative procedure: we pick some point $x$, cut open on that point, pick another point $x$, construct another Hamiltonian of form (\ref{cross1}), cut open there, and keep repeating, until we
 turn $K_1$ into a tree.  
Then, we can find a trivial state on that tree.  Note that while the Hamiltonian of form (\ref{cross1}) may have an increase in the degree $d$ on the ancilla degrees of freedom due to adding the terms $P_{L^a}$ and $P_{R^a}$, we can choose this procedure so that on subsequent steps the complexes $C(L^a)$ and $C(R^a)$ have no intersection with $x$.  That is, when ``cutting" $K_1$ on future steps, we avoid putting the cut on a cell connecting ancilla sites.
Thus, while the degree of the complex may be increased by this procedure (and may be increased on the first step), the degree of the complex can increase by at most a bounded amount over all steps.

However, it is not always possible to find such projectors and such a unitary, as shown by the following toy counter-example.  $L$ and $R$ each consist of a single site.  Each of these sites has a three-dimensional Hilbert space with states
denoted $|1\rangle,|2\rangle,|3\rangle$.  We define $H_{LR}$ to project onto the states where both $L$ and $R$ are in the same basis:
\be
\label{counterexample}
H_{LR}=\sum_{j=1}^3 |j\rangle\langle j| \otimes |j \rangle \langle j|.
\ee  
Suppose the projectors $|j\rangle\langle j|$, for $j=1,2,3$ are all in the interaction algebras of $H_{L^rO}$ and $H_{R^rO}$ on $L^r$ and $R^r$, respectively.
Then, there is no way to pick a Hamiltonian of form Eq~.(\ref{cross1}) such that the given unitary exists.  For example, if we pick $P_{L^a}$ to project onto state $|1\rangle$ and $P_{R^a}$ to project onto state $|2\rangle$, then the state $|3\rangle\otimes |2\rangle$ on the real degrees of freedom would be a ground state of such a Hamiltonian that could be mapped to a ground state of $H$ by a unitary that commutes with the central elements, but the state $|3\rangle\otimes |3\rangle$ could not be mapped in this way, despite being a ground state of Hamiltonian (\ref{cross1}).

However, we claim (and show in the next paragraphs) that this problem never occurs for stabilizer Hamiltonians.
As a result, we can always find a Hamiltonian of the form Eq.~(\ref{cross1}) for each cut such that the Hamltonian fulfills the desired properties: namely, it is a sum of commuting projectors and there is a unitary which on the real and ancilla degrees of freedom on$L,R$ which transforms every ground state of Eq.~(\ref{cross1}) to a ground state of $H$, thus proving theorem \ref{stabtriv}.

Consider a Pauli Hamiltonian $H_{LR}$.  This may have certain nontrivial central elements in its interacton algebra on $L$.  The central elements on $L$ consist of the trivial central element (the identity matrix) and the nontrivial central elements which are products of Pauli operators.  We write generators for the center of the algebra as $C_L^1,C_L^2,...$, choosing $(C_L^a)^2=1$ and choosing all the operators to be independent (independent here means that no product of them is the identity; this implies that all of these $C_L^1,C_L^2,...$ are nontrivial central elements).  Similarly, we use $C_R^1,C_R^2,...$ to denote generators of the center of the interaction algebra on $R$.  If there are $n_L,n_R$ central elements on $L,R$ respectively, then there are $2^{n_L+n_R}$ different subspaces which are different eigenspaces of these operators $C_L{1,2,...},C_R^{1,2,...}$.  The Hamiltonian $H_{LR}$ commutes with these central elements, so we can diagonalize the Hamiltonian in each of these subspaces.  If the Hamiltonian does not have a zero energy state in a given subspace, then the minimum energy in that subspace may be some quite complicated function of the eigenvalues of the central elements.  However, whether or not the Hamiltonian does have a zero energy state in a given subspace can be determined by linear algebra.  The result is that the Hamiltonian has a zero energy state if and only if some set of operator equations of the form
\be
\label{sigmb}
c_L^b c_R^b=\sigma_b
\ee
are obeyed, where $\sigma_b$ is a sign: $\sigma_b=\pm 1$, and $c_L^b$ is some operator that is a product of central elements $C_L^{b_1} C_L^{b_2} ...$ in the interaction algebra on $L$ and $c_R^b$ is some similar operator on $R$.
Assume that a zero energy ground state of $H_{LR}$ does exist. Let some such ground state exist in the state $C_L^b=\tau_L^b,C_R^b=\tau_R^b$ for some functions $\tau_{L,R}^b=\pm 1$.  This state will correspondingly have $c_L^b=\theta_L^b,c_R^b=\theta_R^b$ for some functions $\theta_{L,R}^b=\pm 1$.  Set $P_{R^a}$ to project onto the given eigenspace of the operators $C_R^b$ (i.e., onto the space such that $C_L^b=\tau_L^b$) and set $P_{L^a}$ to project onto the given eigenspace of the operators $C_L^b$.  We claim that this has the desired properties.
Define $c_{L^a}^b$ and $c_{R^a}^b$ in the natural way, being the analogue of the operators $c_L^b,c_R^b$ except acting on the ancilla degrees of freedom.
Then in this eigenspace, $c_{L^a}^b=\theta_L^b$ and $c_{R^a}^b=\theta_L^b$.  The Hamiltonian of Eq.~(\ref{cross1}) will agan commute with all the central elements (operators $C_{L^r}^b,C_{R^r}^b,C_{L^a}^b,C_{R^a}^b$ acting on the real and anclla degrees of freedom, so we can again diagonalize it in each of the $2^{2n_L+2n_R}$ different subspaces.  However, the operators $1-P_{L^a}$ and $1-P_{R^a}$ constrain the choice of central elements on the ancilla degrees of freedom, implying that any zero energy state is in one of the $2^{n_L+n_R}$ different subspaces with given eigenvalues of the central elements on the ancillas.  Applying Eq.~(\ref{sigmb}) to the Hamiltonian $H_{L^rR^a}$ in the given subspace of the ancilla central elements implies that we will have $c_{L^r}^b=\sigma_b \theta_R^b$ and similarly we will have $c_{R^r}^b=\sigma_b \theta_L^b$.  Thus, we will have $c_{L^r}^b c_{R^r}^b=\theta_R^b \theta_L^b=\sigma_b$.  Thus, every zero energy eigenstate of Hamiltonian Eq.~(\ref{cross1}) obeys
\be
\label{sigmbreal}
c_{L^r}^b c_{R^r}^b=\sigma_b,
\ee
and hence we can find a unitary acting on $L,R$ which commutes with the central elements on $L,R$ and which transforms all such zero energy eigenstates of Hamiltonian Eq.~(\ref{cross1}) into zero energy ground states of $H_{LR}$.

We conjecture that it is possible to find trvial ground states for any Hamiltonian with a $1$-localizable interaction complex.  One approach to this is as follows. In the toy example above, we could have added projectors $P_{L^r}$ and $P_{R^r}$ projecting onto central elements of the interaction algebra on the {\it real} degrees of freedom without worrying about ruining the property of $K_2$ being $1$-localizable with given range because $L$ and $R$ consisted of just a single site.  For every example we have tried, in fact, by writing a Hamiltonian of form Eq.~(\ref{cross1}) with also adding additional interaction terms on the real degrees of freedom (added so that each such term has the same support as a pre-existing term), we have been able to find a Hamiltonian with the property that the desired unitary exists.
We conjecture that this is always true.

This problem has an interesting relation with ``topological order": in the toy example, all the central elements were generated by central elements in the interaction algebra on $L$ or $R$ of terms $h_Z$ that appeared in $H_{LR}$ (in this toy example, there is only one term $h_Z$ appearing in $H_{LR}$ with $Z$ consisting of two sites, one in $L$ and one in $R$, so this property follows automatically since the interaction algebra of $H_{LR}$ on $L$ is the same as the interaction algebra of $h_Z$ on $L$).  Whenever this property holds (that the central elements can be generated in this fashion), it is possible to added projectors onto the central elements by adding terms to the Hamiltonian whose support is the same as some pre-existing term (or indeed whose support is smaller than a pre-existing term; in the toy example we add a term whose support is just on $L$ which is a subset of $Z$).
So, we may say that something similar to topological order is present (be aware: this property is different from what we have referred to before as topological order, as it may occur even with a trivial ground state) when central elements occur in the interaction algebra of $H_{LR}$ on $L$ which are not generated by central elements of the interaction algebras of $h_Z$ on $L$ for the $h_Z$ which contribute to $H_{LR}$.  That is, each term $h_Z$ contributing to $H_{LR}$ may or may not have central elements in its interaction algebra on $L$, but there may be central elements in the interaction algebra of $H_{LR}$ on $L$ which are {\it not} generated by these central elements.  To see an example of this phenomenon, consider the following Hamiltonian which is similar to a toric code Hamiltonian with some additional boundary terms.  Let $L$ consist of $3$ sites, called $1,2,3$, and let $R$ also consist of $3$ sites, called $4,5,6$.  Consider the Hamiltonian
\be
H_{LR}=h_{1245}+h_{2356},
\ee
with
\begin{eqnarray}
h_{1245}&=&\sigma_1^z \sigma_2^z \sigma_4^z \sigma_5^z,
\\ \nonumber
h_{2356}&=& \sigma_2^x \sigma_3^x \sigma_5^x \sigma_6^x+\sigma_3^z\sigma_6^z.
\end{eqnarray}
Then, $\sigma_1^z\sigma_2^z\sigma_3^z$ is a central element of the integral algebra of $H_{LR}$ on $L$.  However, the interaction algebra of $h_{1245}$ on $L$ is generated by $\sigma_1^z\sigma_2^z$ and the interaction algebra of $h_{2356}$ on $L$ has no central elements, so the central element of $H_{LR}$ on $L$ is not generated by central elements of the individual terms.
However, in every example we have found, even when this kind of ``topological order" is present, we have been able to add additional projectors to the original Hamiltonian (that is, the Hamiltonian before making the construction of Eq.~(\ref{cross1}) such that the support of the interaction terms is not increased and such that now a Hamiltonian of the form Eq.~(\ref{cross1}) with the desired properties can be constructed.

\section{Hyperfinite Families of Complexes}
The above result suggests that in an attempt to prove the quantum PCP conjecture we should consider interaction complexes which {\it cannot} be turned into $1$-localizable complexes by removing a small fraction of cells.  Here we define such families.

\begin{definition}
Consider a family of $l$-complexes, $K_l(N)$, for $N=1,2,3,...$, where $N$ is the number of $0$-cells in $K_l(N)$.  Such a family is said to be ``$k$-hyperfinite" if
for all $\epsilon>0$, there exists an $R$ such that for all $N$ one can remove at most a fraction
 $\epsilon$ of the $0$-cells $K_l(N)$, while removing also all attached higher cells, such that the resulting complex is $k$-localizable with range $R$.
\end{definition}

To understand this definition, consider by analogy the case of $0$-localizable complexes.  These are complexes that consist of disconnected sets, each of diameter at most $R$.  It is known that there are families of graphs which are not $0$-hyperfinite, despite there being a uniform bound on the number of $1$-cells attached to each $0$-cell.  For example, consider a family of expander graphs with a uniform bound on the degree of the graph.  To disconnect such an expander graph into sets of diameter $R$ small enough that the number of sites in a set of diameter $R$ (which is at most exponential in $R$) is small compared to $N$ requires deleting at least some non-zero fraction $\epsilon_0$ of the edges because each such set has a lower bound on the number of edges leaving it divided by the number of vertices in that set.  Thus, for $\epsilon<\epsilon_0$, the diameter of the sets must grow logarithmically with $N$.

However, the property of being $0$-hyperfinite is stronger than the property of not being a family of expander graphs.  For example, a family of graphs, such that each graph consists of a square lattice with $N/2$ sites and an expander graph with $N/2$ sites, with no connection between the square lattice and the expander graph, is neither $0$-hyperfinite (because it contains the expander graphs) nor a family of expander graphs (because it contains the square lattice).  However, a family of graphs consisting just of square lattices with $N$ sites is $0$-hyperfinite.

\section{Main Result and Discussion}
Our main result is the ability to solve Hamiltonians on a more general class of graphs than considered in Ref.~\onlinecite{bv}. 
Elsewhere\cite{fh}, it will be shown that for any $k$ there exist families of interaction complexes which are not $k$-hyperfinite, even though these families of complexes have a uniform bound on the number of $m+1$-cells attached to every $m$-cell for all $m$; such complexes will be shown to be useful in quantum coding theory and are a natural place to look for proving a quantum PCP conjecture.

Some intuition about these complexes can be thinking of motions of particles and strings.  Intuitively, $0$-localizable complexes can be seen as restricting the motion of particles: consider a $0$-localizable complex $K_1$ which can be mapped to a $0$-complex $K_0$ and consider a particle hopping from $0$-cell to $0$-cell on $K_1$ by following $1$-cells.  The image of the particle's position is always in the same $0$-cell in $K_0$.
At a similar intuitive level, $1$-localizable complexes can be seen as restricting the motion of strings.  Consider the motion of a path on a $1$-localizable complex; for intuitive reasons, let us think about the motion on the covering space which can be mapped to a tree.  One can deform the path by adding ``tendrils" which move out along branches of the tree and return, but the motion is much more restricted than in a plane where a string can sweep out large areas.  This suggests that some algebraic definition of $1$-localizable or $1$-hyperfinite complexes might be possible in terms of a Laplacian for strings.

\subsection{Generalizations}
One problem for future work is to remove the assumption that the terms in the Hamiltonian are commuting.  
Suppose the interaction graph $G$ is a planar square lattice.  Previously, we noted that if holes were punched out of the lattice then it would be $1$-localizable and a trivial ground state could be found for a commuting projector Hamiltonian.  We hope that if larger holes are removed from the lattice, then it will be possible to give low energy witnesses for arbitrary Hamiltonians.  Perhaps, in keeping with the ideas of quantum belief propagation\cite{qbp1,qbp2,qbp3}, it will be possible to approximately describe the thermal state at inverse temperature $\beta$ if the hole circumference is sufficiently large compared to $\beta$ (see also Ref.~\onlinecite{iarad}).  We further hope that such an approach can be extended to any case in which given a graph $G$, the complex obtained by attaching a $2$-cell to every triangle in $G^R$ (for some $R$ sufficiently large compared to $\beta$) is $1$-localizable.
This is a speculative idea for the future.

Another problem for future work is to remove the limitation on the number of interaction terms that can act on a given site.   Our results in subsection \ref{gs1l} did not depend upon this number.  However, if the number of interaction terms that acts on a given site is large, then the degree of the interaction graph is large and hence a unitary of bounded range may act on a large number of sites, making it harder to compute the energy of a trivial state.  If the degree of the interaction graph grew sufficiently rapidly with $N$, then we would not even be able to approximately write down an arbitrary unitary of bounded range using resources that are polynomial in $N$.  This is why we would like a different approach to deal with this case.
However, suppose that a given site $0$ couples to many other sites $1$,$2$,...,$n$, for some $n>>1$.  Intuitively, given that the Hilbert space dimension of site $0$ is bounded, it is not possible for site $0$ to be strongly entangled with all of those other sites and so a product or mean-field approximation becomes useful.  
As an example, consider a system of spin-$1/2$ spins and suppose the Hamiltonian includes $\sum_{i=1}^n \vec S_0 \cdot \vec S_i$, where $\vec S_i$ is the vector of spin operators on site $i$.  Then, at a small cost in energy density, we can fix the spin $0$ to point in a given direction, considering a state which is a product state of spin $0$ with the rest of the system.  Perhaps using tools from monogamy of entanglement, it will be possible to deal with this case as well.

{\it Acknowledgments---} I thank M. Freedman, K. Walker, and Z. Wang for many very useful discussions, especially in terms of formulating this problem in terms of simplicial complexes.  I thank D. Poulin for useful discussions on quantum belief propagation.
This material is based upon work supported in part by the National Science Foundation under Grant No. 1066293 and the hospitality of the Aspen Center for Physics.

\appendix
\section{Shields and Relation to Cover}
Here, we relate the property of $K_2$ being $1$-localizable to properties of covers of $K_2$.  We show that if a complex is $1$-localizable then the complex has a cover which as a continuous mapping $g$ to a $1$-complex $T_1$ which is a tree, such that the map has pre-images with bounded diameter (the relation between the diameter of the pre-image under $g$ to the diameter under the pre-image under $f$ is discussed later).
We conjecture that the converse is true but do not have a general proof for arbitrary covers.

Our reason for the interest in studying this is that it naturally relates to ideas of so-called ``quantum belief propagation", as solving a problem on a cover of the original graph is reminiscent to the iterative nature of belief propagation equations.  The idea of a ``shield" developed below has some relation to the Markov shield in Ref.\onlinecite{qbp3}, and helps explain the decomposition of Hamiltonians used in section \ref{gs1l}.  We note, however, that nothing in this appendix is necessary for other parts of the paper.

\begin{figure}
\includegraphics[width=1.9in]{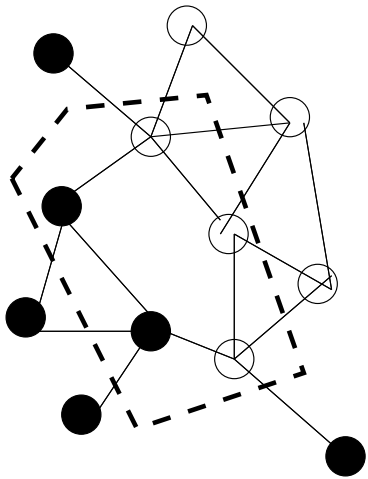}
\caption{Illustration of shield.  $X$ is the set surrounded by a dashed line.  Circles and lines correspond to $0$- and $1$-cells, respectively.  Each triangle is assumed to have a $2$-cell attached.  The set $X$ has $5$ shields.  One shield is shown using the open circles: the $0$-cells with open circles are all in one shield with the three circles inside the dashed line being in the interior shield and the other three being in the exterior shield.  Note that two of the $0$-cells with open circles (the top-most and bottom-most such $0$-cells in $X$) are each also in one other shield.}
\label{shield}
\end{figure}

First we define the shields of a set:
\begin{definition}
Given a complex $K_2$ and a set $X$ of $0$-cells in $K_2$, let $Y$ be the set of $0$-cells at distance $1$ from $X$.
Define a graph $H$, with vertex set being the set of ordered pairs $(i,j)$ for which $i$ is a $0$-cell in $X$ and $j$ is a $0$-cell in $Y$ and
$i$ and $j$ are connected by an $1$-cell in $G$.
Let there be an edge in $H$ connecting vertex $(i,j)$ to $(k,l)$ if $i=k$ and there is a $2$-cell containing the $0$-cells $i,j,l$ or if $j=l$ and there is a $2$-cell containing the $0$-cells $i,j,k$.

Define the ``shields of $X$" as follows.  Pick any $0$-cell $i$ in $X$ and any other $0$-cell $j$ in $Y$.  Let
$S(i,j)$ be the set of vertices in $H$ which are connected, by a path in $H$, to the vertex $(i,j)$ in $H$.
The sets of shields of $X$ is the set of all sets $S$ such that $S=S(i,j)$ for some $i,j$.
In a slight abuse of notation, we say that a $0$-cell $i$ is in a shield $S$ if $(i,j)$ is in $S$ for some $j$ or $(j,i)$ is in $S$ for some $j$.

Given a set $S$ which is a shield of $X$, we say that the set of $0$-cells $j$ such that $(i,j)$ is in $S$ for some $i$ is an ``exterior shield" of $X$ and the
set of $0$-cells $i$ such that $(i,j)$ is in $S$ for some $j$ is an ``interior shield" of $X$.
\end{definition}
Fig.~\ref{shield} shows an example of shields.  Note that the same $0$-cell can be in more than one shield.

Given a set $X$, we can write the Hamiltonian as a sum of commuting terms as
$H=H_X+\sum_s H_{X,s}+H_{\overline X}$, where $H_X$ is supported on the set $X$, $H_{\overline X}$ is supported on the complement of $X$ and the sum $s$ ranges over the shields of set $X$.  Each term $H_{X,s}$ is supported on a given shield (it is supported both on the interior and exterior shield of that shield).  The interaction algebra of $H_{X,s}$ on $X$ commutes with the interaction algebra of $H_{X,s'}$ on $X$ for $s\neq s'$.  This decomposition into different $H_{X,s}$ is the decomposition that we used in defining the $H_0(i_j)$ in section ~\ref{gs1l}: each $i_j$ corresponds to a distinct shield $s$.

Shields are also related to the concept of ``Markov shield" in Ref.~\onlinecite{qbp3}.  For commuting projector Hamiltonians with unique ground states, in order to saturate the strong subadditivity inequality Eq. (1) of Ref.~\onlinecite{qbp3} for the reduced density matrices of the ground state when adding a site $k$ in the notation of that paper, it suffices that the Markov shield be an interior shield of the set of sites $\{1,...,k-1\}$ and $k$ must be a site in the corresponding exterior shield.   Indeed, if site $k$ being added is replaced with a {\it set} of sites being added, and if that set of sites is an exterior shield, then it suffices that the Markov shield be the corresponding interior shield to saturate the inequality.  To see this, note that the interaction terms with support on the interior shield can be written as a sum of two terms; one term contains the interactions supported on the shield (supported on both the interior and the corresponding exterior shields), and the other term contains everything else.  The interaction algebras of these terms on the interior shield commute.
The Hilbert space on the interior shield can be decomposed as before into a sum of products of Hilbert spaces.  Assuming the Hamiltonian has a unique ground state, then the density matrix is supported only one of these terms in the direct sum.  Then, this term in the direct sum decomposes into a product of two Hilbert spaces, and the interactions supported on the shield involve only one of these spaces while the other interactions involve the other space.

In many cases, one can saturate the inequality with a smaller Markov shield; for example, for the toric code Hamiltonian, if the sets $\{1,...,k-1\}$ and $\{1,...,k\}$ have the same topology, then it suffices that the Markov shield contain only the sites in $\{1,...,k-1\}$ that interact with site $k$.  However, this is not always sufficient and if the topology changes then one may need to take the Markov shield equal to an interior shield to saturate the inequality for this Hamiltonian.

Shields also have an interesting relation to junction trees.  A junction tree decomposition of a graph $G$ is a tree graph $T$, with each vertex of $T$ being associated with a set of vertices in $G$.  The union of these sets of vertices is the set of vertices in $G$.  For every edge $(i,j)$ in $G$ there is a vertex in $T$ whose associated set includes $i$ and $j$.  Finally, the set of vertices in $T$ whose associated sets contain any given vertex $i$ of $G$ is a connected set.
Consider any
 set $X$ associated to any vertex $v$ in $T$.  Any vertex $i$ in $G$ that is a first neighbor of $X$ must be in a set associated to some vertex $z$ neighboring $v$ (there is an edge between $i$ and some vertex $j\in X$, so $i$ and $j$ are both in some set $Y$ associated to a vertex $w$ of $T$; $i$ is not in $X$ so $v \neq w$; let $z$ be the neighbor of $x$ on the shortest path from $v$ to $w$; since the set of vertices containing $i$ is connected, $i$ must be in $z$).  
However, given any $i$ which is a
first neighbor of $X$ and any $k$ which is a first neighbor of $X$, if there is an edge from $i$ to $k$ in $G$ then it is not possible that $i$ is in a set associated to some given neighbor $z$ of $v$ and $k$ is in a set which is associated to some other neighbor $z'\neq z$ (if there is such an edge from $i$ to $k$, then there is some set containing both $i$ and $k$; however, since $X$ contains neither $i$ not $k$, this contradicts the assumptions that the set of vertices in $T$ whose associated sets contain any given vertex of $G$ is connected).
So, if we attach a $2$-cell to every triangle of $G$ to define a complex, each shield of any set $X$ associated to any vertex $v$ in $T$ contains only vertices in $X$ and in some given set associated to a neighbor of $v$.

Finally, shields are useful in how they relate to covers of complexes; roughly, one uses the shields to define various transition functions to construct a cover of the complex such that the cover can be coarse-grained into a tree.  We give two definitions of certain families of complexes, one involving covers and one involving shields.  Both of these definition have natural interpretations in terms of quantum belief propagation.  We relate these definitions to each other.  Finally, we relate these definitions to our previous definition of $1$-localizable complexes.

We define
\begin{definition}
A complex $K_2$ is {\bf cover $1$-localizable} with range $R$ if there is a complex $\tilde K_2$ that is a cover of $K_2$ such that there exists a map $f$
from $\tilde K_2$ to a $1$-complex $T$ that is a tree, such that the preimage of any point under $f$ has diameter at most $R$.
\end{definition}
and also
\begin{definition}
\label{defn2}
A complex $K_2$ is set $1$-localizable with range $R$ if it is possible to find sets of $0$-cells $C_a$, for $1 \leq a \leq n$ for some $n$ (possibly infinite) such that
\begin{itemize}
\item[{\bf 1}:] The diameter of each set $C_a$ is at most $R$
\item[{\bf 2}:] Every $0$-cell is in at least one such set
\item[{\bf 3}:] For every set $C_a$, and for every shield $S$ of $C_a$, there is some other set $C_b$ and some shield $T$ of $C_b$ such that $S$ is equal to $T$ up to transposition of entries (i.e., $T$ is equal to the set of all $(i,j)$ such that $(j,i) \in S$).  In this case we say that $C_a$ and $C_b$ ``are neighbors".
\end{itemize}
\end{definition}

Before relating these definitions, we relate the cover definition to a definition in terms of coarse-graining.
  We define coarse-graining a complex analogously to defining coarse-graining a graph:
\begin{definition}
Given a complex $K_2$, we define a coarse-grained graph $K_2'$ as follows.  Let $C_1,C_2...$ be sets of $0$-cells of $K_2$, called these ``clusters".  Let these clusters be disjoint, and let each $0$-cell of $K_2'$ be in one of the clusters.  Then, the coarse-grained complex $K_2'$ has one $0$-cell corresponding to each cluster, and there is an $1$-cell attached to $0$-cells $i$ and $j$ $(i,j)$ in $K_2'$ if and only if there is a $0$-cell $v \in C_i$ and a $0$-cell $w\in C_j$ such that there is a $1$-cell attached to those $0$-cells.  Further, the coarse-grained complex has a $2$-cell attached to $1$-cells attached to $0$-cells $i,j,k$ if and only if there is a $2$-cell in $K_2$ attached to $1$-cells attached to $0$-cells such that the three $0$-cells are in $C_i,C_j,C_k$ respectively with $i \neq j \neq k \neq i$.
\end{definition}

\begin{lemma}
If a complex $K_2$ is cover $1$-localizable with range $R$ then some cover of $K_2$ it can be coarse-grained into a complex with no $2$-cells, the number of $0$-cells in each cluster bounded by some function of $R$ and the degree $d$.  Conversely, if some cover of $K_2$ can be coarse-grained into a complex with no $2$-cells, then $K_2$ is cover $1$-localizable with a range $R$ that bounded by some function of the degree $d$ and the maximum number of $0$-cells in each cluster.
\begin{proof}
The converse direction is immediate: the coarse-grained cluster can be obtained from $K_2$ by some continuous mapping, mapping each $1$-cells connecting $0$-cells in a cluster to a point.  So, there is a continuous mapping from some cover of $K_2$ to a $1$-complex, so there is a continuous mapping of a cover of $K_2$ to a tree since the universal cover of the $1$-complex is a tree.

To show that converse, assume that $K_2$ is cover $1$-localizable, so some cover $\tilde K_2$ can be mapped to a tree $T_1$.   
We can assume (as above) that $f$ is good, at the cost of increasing $R$, so each $0$-cell of $\tilde K_2$ is mapped to
a $0$-cell of $T_1$.  Each $1$-cell is mapped to a path of length at most $l_{max}$ as discussed previously.  Hence, $H$ is a sub-graph of $T^{l_{max}}$, for some tree graph $T$.  We have described above how to coarse-grain such graphs into a triangle-free graph.
\end{proof}
\end{lemma}

It is worth noting that we only require that $\tilde K_2$ being {\it a} cover of $K_2$, rather being the universal cover.  To see why we chose this, imagine the following complex.  Consider, for example a triangulation of a very long thin torus: that is, take the length in one dimension to be of order unity while the length in the other direction is of order $N$.  Then, the universal cover is a triangulation of the plane and has no map to a tree, but there is a cover (namely, an infinitely long thin cylinder) which can be coarse-grained into a tree (indeed, coarse-grained into a line).

\begin{lemma}
Any cover $1$-localizable complex with range $R$ and degree $d$ is also set $1$-localizable with range bounded by a function of $R,d$.
\begin{proof}
By above, there is a cover of the complex that can be coarse-grained to a tree.  Let the set of $C_a$ be the set of clusters in the coarse-graining.
This fulfills conditions {\bf 1,2} immediately.

Since the coarse-graining is to a tree, for each $C_a$, for each shield of $C_a$ there is some $C_b$ such that the shield contains only pairs $(i,j)$ with $i \in C_a$ and $j \in C_b$.  Thus, the shield is also a shield of $C_b$ up to a transposition of entries.
\end{proof}
\end{lemma}

\begin{lemma}
Any set $1$-localizable complex with range $R$ is a cover $1$-localizable complex with range $R$.
\begin{proof}
 Assume without loss of generality that $K_2$ is connected (if it is not, repeat this procedure on each connected component).

We construct the cover as follows.  Pick any set $C_a$.
The $0$-cells in $\tilde K_2$ are labelled by $i$, where $i$ is a $0$-cell in $K_2$, and by a finite sequence $a_1,a_2,...,a_n$, where $a_1=a$ and $C_{a_{n}}$ contains $i$ and for each $i$, $C_{i+1}$ is a neighbor (as in definition (\ref{defn2}) of $C_i$ and where $C_{i+2} \neq C_i$ for any $i$.  That is, one labels the $0$-cells
by a $0$-cell in $K_2$ as well as a non-contractible path in the graph whose vertices are the sets $C_a$ with edges connecting any two $C_a,C_b$ which are neighbors.  We write such a label as $(i,P)$, where $P$ is the path.  The covering map maps each such $(i,P)$ to $0$-cell $i$.

We attach a $1$-cell between any two $0$-cells $(i,P)$ and $(j,P')$ if there is a $1$-cell in $K_2$ attached to $i,j$ and if either, $P=P'$ (in which case both $0$-cells are in the same $C_a$) or $P$ is the same sequence as $P'$ except for either adding or removing entry at the end so that either $P=a_1,...,a_n$ and $P'=a_1,...,a_{n+1}$ or $P=a_1,...,a_{n+1}$ and $P'=a_1,...,a_n$.  The covering map maps each such $1$-cell to the $1$-cell attached to $i,j$.

We attach  $2$-cells to three $1$-cells in $\tilde K_2$ as follows.  Suppose the three such $1$-cells are attached to three $0$-cells $(i,P), (j,P'), (k,P'')$.
Then, we attach the $2$-cell if a $2$-cell is attached to the corresponding $1$-cells in $K_2$ and either $P=P'=P''$ or two of three paths are the same (i.e., either $P=P'$ and $P'' \neq P$ or either of the two other possibilities) and the third distinct path differs only by either adding or removing a single entry at the end.  Note that by the definition of the shield, we never have three $1$-cells attached to three $0$-cells $(i,P),(j,P'),(k,P'')$, with $P,P',P''$ all distinct and with there being a $2$-cell in $K_2$ attached to the image of those $1$-cells.

Given that this is a covering map, the map to a tree is as follows: map all $0$-cells with a given path $P$ to a single vertex in $T$, calling that vertex $P$.  Map all $1$-cells and $2$-cells such that all $0$-cells in that cell are contained in some given $P$ to the same vertex $P$ also.  Map all $1$-cells and $2$-cells such that there are two distinct paths, $P,P'$ containing $0$-cells in that $1$- or $2$-cell to the edge connecting $P$ to $P'$ (this map can be made continuous in the natural way, mapping points in the cell closer to the $0$-cell in $P$ to points in the edge close to $P$).
Note that in order to get a tree it was essential that we never have a $2$-cell containing $0$-cells in three distinct $P,P',P''$. 
\end{proof}
\end{lemma}

Now consider the relation of these definitions to the definition of a $1$-localizable complex.
It is immediate that any $1$-localizable complex with range $R$ is cover $1$-localizable with range $R$.
We conjecture that the converse is true, namely that
any cover $1$-localizable complex $K_2$ with range $R$ and degree $d$ is also $1$-localizable with range depending only upon $R,d$.  However, we do not give a proof of this statement, though in every example we have considered this conjecture holds.

\end{document}